\documentclass[aps,floatfix,superscriptaddress,nofootinbib,longbibliography,preprintnumbers,11pt,tightenlines]{revtex4-1}
\usepackage[paperwidth=210mm,paperheight=297mm,centering,hmargin=2.3cm,vmargin=2.3cm]{geometry}
\usepackage{amssymb,amsthm,amsfonts,braket,amsmath,graphicx,color,dsfont, mathrsfs}
\usepackage{array}
\newcolumntype{C}[1]{>{\centering\let\newline\\\arraybackslash\hspace{2pt}}m{#1}}
\newcolumntype{R}[1]{>{\RaggedRight\let\newline\\\arraybackslash\hspace{2pt}}m{#1}}
\usepackage{bm}
\usepackage{enumerate}
\usepackage{url}
\usepackage{xcolor}
\usepackage{multirow}
\usepackage[colorlinks=true,citecolor=blue,linkcolor=blue
]{hyperref}
\usepackage{float}
\usepackage[normalem]{ulem}
\newcommand{\goldieotimes}{
  \mathop{\mathchoice{\textstyle\bigotimes}{\bigotimes}{\bigotimes}{\bigotimes}}
}
\newcommand{\jmax}{j_\mathrm{max}}
\begin{document}

\dimen\footins=5\baselineskip\relax

\title{Eigenstate Thermalization in 1+1-Dimensional SU(2) Lattice Gauge Theory Coupled with Dynamical Fermions}

\author{Diptarka Das}
\email{didas@iitk.ac.in}
\affiliation{Indian Institute of Technology - Kanpur, 208016, U.P, India}

\author{Lukas Ebner}
\email{lukas.ebner@mpq.mpg.de}
\thanks{corresponding author and leading contributor to numerical results in this manuscript.}
\affiliation{Institut f\"ur Theoretische Physik, Universit at Regensburg, D-93040 Regensburg, Germany}
\affiliation{Max Planck Institute of Quantum Optics, 85748 Garching, Germany}
\affiliation{Munich Center for Quantum Science and Technology (MCQST), 80799 Munich, Germany}

\author{Saurabh V. Kadam}
\email{ksaurabh@uw.edu}
\affiliation{
InQubator for Quantum Simulation (IQuS), Department of Physics, University of Washington, Seattle, WA 98195, USA}

\author{Indrakshi Raychowdhury}
\email{indrakshir@goa.bits-pilani.ac.in}
\affiliation{Department of Physics, Birla Institute of Technology and Science Pilani,
K K Birla Goa Campus, Zuarinagar, Goa 403726, India}\affiliation{Center for Research in Quantum Information and Technology, Birla Institute of Technology and Science Pilani, Zuarinagar, Goa 403726, India}

\author{Andreas Sch\"afer}
\email{andreas.schaefer@physik.uni-r.de}
\affiliation{Institut f\"ur Theoretische Physik, Universit at Regensburg, D-93040 Regensburg, Germany}

\author{Xiaojun Yao}
\email{xjyao@uw.edu}
\affiliation{
InQubator for Quantum Simulation (IQuS), Department of Physics, University of Washington, Seattle, WA 98195, USA}

\date{\today}
\preprint{IQuS@UW-21-108}

\begin{abstract}
We test the eigenstate thermalization hypothesis (ETH) in 1+1-dimensional SU(2) lattice gauge theory (LGT) with one flavor of dynamical fermions. 
Using the loop-string-hadron framework of the LGT with a bosonic cut-off, we exactly diagonalize the Hamiltonian for finite size systems and calculate matrix elements (MEs) in the eigenbasis for both local and non-local operators.
We analyze different indicators to identify the parameter space for quantum chaos at finite lattice sizes and investigate how the ETH behavior emerges in both the diagonal and off-diagonal MEs.  
Our investigations allow us to study various time scales of thermalization and the emergence of random matrix behavior, and highlight the interplays of the several diagnostics with each other. 
Furthermore, from the off-diagonal MEs, we extract a smooth function that is closely related to the spectral function for both local and non-local operators. We find numerical evidence of the spectral gap and the memory peak in the non-local operator case.
Finally, we investigate aspects of subsystem ETH in the lattice gauge theory and identify certain features in the subsystem reduced density matrix that are unique to gauge theories. 
\end{abstract}

\maketitle

\tableofcontents

\section{Introduction
\label{sec: Introduction}
}

Understanding how an isolated quantum system thermalizes or approaches a thermal equilibrium is far from fully understood for strongly correlated systems. 
Since the early 90s, a series of theoretical developments have established and conjectured a deep connection between the thermalization properties of a quantum system and matrix elements (MEs) of local observables in the energy eigenbasis, highlighted in the formulation of the eigenstate thermalization hypothesis (ETH)~\cite{Deutsch:1991msp,Srednicki:1994mfb,rigol2008thermalization}. 
Physically, it states that a highly excited pure quantum state looks like a thermal state when probed locally. The ETH has been widely tested for non-integrable spin systems~\cite{Santos:2010iji} and has been extremely successful in explaining statistical features of various condensed matter, and atomic systems, as well as optical physics~\cite{Chandran:2016vow, Jansen_2019, Lu:2017tbo, Bao:2019bjp, Langen:2014saa}.
It is also consistent with conformal properties of highly excited states expected at quantum critical points~\cite{Lashkari:2016vgj, Basu:2017kzo}. 
More recently, technological progress has allowed us to perform highly controllable quantum simulations to probe the thermalization properties of strongly correlated quantum systems~\cite{Halimeh:2025vvp,Bauer:2023qgm,Somhorst:2021tuq,Mueller:2024mmk,Farrell:2024mgu,Andersen:2024aob,Perrin:2024kub,Zhou:2021kdl,deJong:2021wsd,Than:2024zaj,Luo:2025qlg,Davoudi:2022uzo}.

Similar studies of thermalization for gauge theories that describe the fundamental forces in nature in terms of interacting quantum fields are needed to understand physics from the microscopic scales of particle collisions to macroscopic phenomena in cosmology and astrophysics.
Specifically, for quantum chromodynamics (QCD), which is the non-Abelian gauge theory that describes the strong force between quarks and gluons, thermalization plays a key role in interpreting the experimental data measured in relativistic heavy ion collisions~\cite{Fries:2008vp,Berges:2020fwq}.
Previous phenomenological studies for heavy ion collisions used kinetic theory~\cite{Baier:2000sb,Kurkela:2018wud,Brewer:2022vkq,Rajagopal:2025nca}, classical Yang-Mills simulation~\cite{Dusling:2010rm,Schenke:2012wb,Kurkela:2012hp}, and holographic correspondence~\cite{Chesler:2008hg,Balasubramanian:2010ce,Ishii:2015gia} to study the rapid thermalization, or more precisely, hydrodynamization, in the initial stages of the collision.
Although great insights have been gained from these studies, their application to phenomenology and explanation of the experimental data may be limited due to unknown systematic uncertainties such as those caused by quantum effects. 
To describe quantum field theoretical properties of QCD observables in these collisions requires nonperturbative methods, as QCD becomes strongly interacting at the relevant energy scales.
Traditionally, the lattice formulation of QCD with Euclidean time~\cite{Wilson:1974sk} in conjunction with Monte-Carlo methods has been used for nonperturbative calculations~\cite{Guenther:2020jwe,Aarts:2023vsf}.
Despite its successes, the Euclidean nature of the formulation limits its applicability for accessing non-equilibrium dynamics.
Alternatively, Hamiltonian lattice formulations of QCD with Minkowski time have been proposed in the past, for example the Kogut-Susskind (KS) formulation~\cite{Kogut:1974ag}, as another nonperturbative tool.
However, limited computational resources prohibited its practical usage due to the exponential scaling of the Hilbert space with lattice size.
Nonetheless, novel computational tools like quantum computing, and tensor and neural network methods have revived the interest in using Hamiltonian simulations for studying lattice gauge theories (LGTs)~\cite{Cataldi:2023xki,Magnifico:2024eiy,Banuls:2019rao,Banuls:2019bmf,Belyansky:2023rgh,Papaefstathiou:2024zsu,Mathew:2025fim,Hayata:2023pkw,Itou:2024psm,Lin:2024eiz,Anderson:2024kfj, Yu:2025czp, Davoudi:2022xmb, Shaw:2020udc, Byrnes:2005qx, Lamm:2019bik, Rhodes:2024zbr, Balaji:2025afl}.
Most of these studies focused on theories simpler than QCD, since the resources needed to simulate QCD are still beyond the current computational capabilities.

Modern high performance computing clusters have enabled us to study LGTs by exactly diagonalizing large matrices, which may act as a benchmarking tool for novel computational methods.
Recent exact diagonalization studies performed by some of the authors systematically demonstrated quantum ergodicity and the ETH behavior for an SU(2) pure gauge theory in two spatial dimensions with continuous time (2+1D) on a chain of up to 19 plaquettes and a honeycomb lattice~\cite{Yao:2023pht,Ebner:2023ixq}.
In these calculations, the infinite dimensional local Hilbert space of gauge bosons was regulated via a truncation scheme to make the Hilbert space finite dimensional.
Various entanglement properties have also been calculated, including the area/volume law of entanglement entropies, the absence of quantum many-body scars~\cite{Ebner:2024mee}, entanglement production in real-time evolution, and the comparison of the entanglement spectrum with the spectrum of the modular Hamiltonian~\cite{Ebner:2024qtu}.
Furthermore, emergent hydrodynamic behavior has been shown in these lattice systems~\cite{Turro:2025sec}. 
These studies were performed without dynamical fermions and continuum limit extrapolations.
While the success of ETH in general for varied systems and the just listed properties of SU(2) pure gauge theory strongly suggest that ETH does also hold for QCD, for such a fundamentally important property one would prefer to have positive evidence that the inclusion of fermions does not invalidate the observed agreement.
Recently, claims have been made about the existence of quantum many-body scars in lower-dimensional gauge theories with dynamical fermions~\cite{Halimeh:2022rwu, PhysRevResearch.7.013322}, which were absent in the pure gauge case~\cite{Ebner:2024mee}.
This hints that presence of dynamical fermions coupled to gauge fields may alter their thermalization behavior significantly, which strongly motivates systematic tests of quantum ergodicity and the ETH in such theories.

In this work, we test quantum ergodicity and the ETH for 1+1D SU(2) LGT coupled with dynamical fermions.
We use the loop-string-hadron (LSH) framework~\cite{Raychowdhury:2019iki, Kadam:2022ipf, Kadam:2024ifg}, which is equivalent to the KS formulation, for constructing the Hilbert space and operator MEs in this theory.
The advantage of the LSH formulation for Hamiltonian simulation lies in its reduced computational complexity 
as demonstrated for the 1+1D SU(2) LGT in Ref.~\cite{Davoudi:2020yln}.
We utilize this for efficient construction of the gauge-invariant Hilbert space of this LGT, and perform the desired tests by exactly diagonalizing the Hamiltonian for a reasonably large physical Hilbert space.

The organization of this manuscript is as follows: In Sec.~\ref{sec: theoretical background}, we provide theoretical background for the ETH, review the KS formulation for the LGT, and lay out the LSH framework. 
Then, we show various ETH test results in Sec.~\ref{sec: Results}, including level statistics of the eigenspectrum, diagonal and off-diagonal MEs of physical operators, and random matrix behavior. Furthermore, we study some prospects of the subsystem ETH that are unique to gauge theories in Sec.~\ref{sec: Garrison-Grover}.
Finally, we summarize and draw conclusions in Sec.~\ref{sec: conclusions}.
 
\section{Theoretical background
\label{sec: theoretical background}
}
In the following, we review the ETH ansatz for physical operators and the features of its components in Sec.~\ref{subsec: ETH}, and then, summarize the LSH formulation of the 1+1D SU(2) LGT with dynamical fermions in Sec.~\ref{subsec: SU2 LGT}.

\subsection{Eigenstate thermalization hypothesis
\label{subsec: ETH}
}

The macroscopic properties of an isolated and highly excited quantum system are successfully predicted by the framework of statistical mechanics.
Nonetheless, their relation to the microscopic dynamics is still not completely understood.
Early efforts in explaining thermalization from microscopic principles led to some key insights.
This included the emphasis on physical observables rather than the wavefunction~\cite{vonNeumann:2010guc}, and conjectures that relate integrability and quantum chaos to level statistics of quantum systems~\cite{berry1977level, Bohigas:1983er, Dyson:1962es, 1962JMP.....3..157D, Dyson:1962oir}.
It is known that many properties of the energy levels of ergodic systems show universal properties, which only depend on the symmetries of the system. The simplest such systems are random matrix ensembles. 
Bohigas, Giannoni, and Schmit (BGS) conjectured that the energy spectrum of all classically chaotic systems that are time-reversal invariant should follow the prediction of a Gaussian orthogonal ensemble (GOE)~\cite{Bohigas:1983er}. 
Thus, whether the level statistics of a quantum system can be described by those of random matrices is often used as an indicator for quantum chaos.
Despite these progresses, the question of how the macroscopic properties at thermal equilibrium emerge in real-time dynamics remains unanswered. Hence, it is worth investigating complicated models in this context.

For local and few-body observables, the ETH offers a plausible explanation in relating microscopic features with macroscopic properties.
It reconciles two pictures of thermalization in a quantum mechanical system at a given energy $E$: i) long-time average of an observable in any initial state with energy close to $E$ and ii) the observable value obtained from an average over a microcanonical ensemble with the ensemble average energy $E$. 
The ETH achieves this by endowing operator MEs of a Hermitian operator $O$ in the energy eigenbasis with the following structure~\cite{Deutsch:1991msp, Srednicki:1994mfb}:
\begin{align}
    \langle E_a | O | E_b \rangle &= O_{\rm mc}(\bar{E})\, \delta_{ab} + e^{-S(\bar{E})/2} \,f_O (\bar{E},\omega)\, R_{ab}, 
    \label{eq: ETH operator ansatz}
\end{align}
where the energy eigenstates are labeled by their energy eigenvalues $E_a$ and $E_b$, $\bar{E}$ is the average energy given by $\bar{E} = (E_a + E_b)/2$, and $\omega = E_a - E_b $ is the difference between the two energies involved.
The dominant part of the diagonal elements, $O_{\rm mc}(\bar{E})$, is a smooth function of $\bar{E}$ that is identical to the expectation value of the microcanonical ensemble at $\bar{E}$.
It is related to its thermal value at a temperature $T$, such that, $T dS = dE$, where the variations in the thermodynamic entropy $S(E)$ are evaluated at the average energy $\bar{E}$.

The off-diagonal elements carry the dynamical information as the time dependence of observables comes from $e^{-i ( E_a - E_b ) t}$ weighted by the elements $\langle E_a | O | E_b \rangle$. 
Thus, they truly control the success or failure to thermalize in real-time dynamics.
The exponential suppression in the correction to the microcanonical expectation value in Eq.~\eqref{eq: ETH operator ansatz} contributes to the extensive entropic suppression of the variance of the observable, which is a necessary condition for thermalization.
The function, $f_O(\bar{E},\omega)$ is a smooth function of its arguments that satisfies $f_O(\bar{E},\omega)= f_O(\bar{E},-\omega)$ for Hermitian Hamiltonians, and the MEs $R_{ab}$ are rapidly varying numbers with unit variance for $a \neq b$ and variance two for $a=b$ and zero mean.
It is expected that the matrix $R$ becomes a Gaussian random matrix when $\omega$ is small, which renders the ETH ansatz to reproduce the predictions of the random matrix theory (RMT) for states in a very narrow energy window.
This also constrains the asymptotic behavior of $f_O(\bar{E},\omega)$ in the $\omega\to0$ limit to approach a constant value.
The energy scale for which ETH reduces to RMT is denoted by $\Delta E_{\rm RMT}$, and the matrix $R$ remains random as long as $\omega \leq \Delta E_{\rm RMT}$.
Constraints from transport studies bound the $\Delta E_{\rm RMT}$ scale to be far below the earlier expected scale given by the Thouless time~\cite{Dymarsky:2018ccu, Richter:2020bkf}.
Finally, $f(\bar{E},\omega)$ is bounded from above by $e^{-\omega/4T}$ at large $\omega$ as a consequence of the analyticity of the thermal two point correlation function~\cite{Murthy:2019fgs}.

The ETH is generally conjectured to hold for many physical observables that fall within the scope of statistical mechanics, including extended few-body operators~\cite{Garrison:2015lva,Dymarsky:2016ntg}, generally referred as the subsystem ETH.
Even though the ETH is formulated in terms of the energy eigenstates, it is valid for states that are expected to thermalize, i.e., those states far away from the ends of the eigenenergy spectrum, where the gap is ${\cal O}(e^{-S(E)})$, i.e., exponentially small between neighboring energy levels~\cite{DAlessio:2015qtq}.
For such a dense region in the eigenenergy spectrum, the eigenvectors within a narrow energy window are physically indistinguishable, which suggests that computing $\langle E | O | E \rangle$ for any state $| E \rangle$ in that energy window would result in a value that is extremely close to the microcanonical average.
The large number of states involved in the microcanonical average also implies that the statistical fluctuations are entropically suppressed because the average separation in energy levels is proportional to $e^{-S(E)}$.
Moreover, since the entropy grows extensively with system size, so does the exponential suppression.
These are statements on the statistics of the diagonal element in Eq.~\eqref{eq: ETH operator ansatz} and the very existence of the diagonal smooth function $O_{\rm mc}(\bar{E})$.
This also leads to the long-time average of observables for an initial state $\ket{\psi}$ to agree with their microcanonical value up to small sub-extensive corrections~\cite{DAlessio:2015qtq}, provided that the degeneracies in the energy spectrum have been removed by lifting all the symmetries.

In this work, we test the ETH ansatz for operators in an SU(2) LGT in 1+1D with dynamical fermions, formulated in the LSH framework.
We now proceed with reviewing the Hamiltonian and the Hilbert space structure within the LSH formulation in the next subsection.

\subsection{SU(2) lattice gauge theory in 1+1D
\label{subsec: SU2 LGT}
}
We consider here a non-Abelian gauge theory with gauge group SU(2) in 1+1D with dynamical fermions.
The Hamiltonian framework of such a LGT was originally proposed by Kogut and Susskind (KS)~\cite{Kogut:1974ag}.
Since then, several different formulations have been proposed over the years, see Refs.~\cite{Banuls:2017ena, Raychowdhury:2019iki, Kadam:2022ipf, Kadam:2024ifg, Chandrasekharan:1996ih, Zohar:2014qma,Zohar:2018cwb,Zohar:2019ygc, Alexandru:2023qzd, DAndrea:2023qnr, Zache:2023dko, Wiese:2021djl, Bergner:2024qjl, Grabowska:2024emw, Ciavarella:2024fzw, Burbano:2024uvn, Illa:2025dou,Yao:2025uxz,Hartung:2022hoz,Fontana:2024rux,Jakobs:2025rvz,mirandariaza2025renormalizeddualbasisscalable,jakobs2025comprehensivestresstesttruncated}. In this section, we first briefly discuss the KS Hamiltonian and the Hilbert space construction for the SU(2) LGT in Sec.~\ref{subsubsec: KS formulation}.
In Sec.~\ref{subsubsec: LSH formulation}, we review the LSH formulation that was first proposed in Ref.~\cite{Raychowdhury:2019iki} and is used for the analysis in Sec.~\ref{sec: Results}.
In order to reach larger system sizes with given classical computation resources, we have identified various symmetry sectors associated with the symmetry transformations of this theory that will be laid out in Sec.~\ref{subsubsec: symmetries}.

\subsubsection{The Kogut-Susskind formulation
\label{subsubsec: KS formulation}
}
The KS framework considers continuous time and a discretized space with the lattice spacing $a$, which we have set to one throughout this work. In other words, all quantities are expressed in units of $a$ or $a^{-1}$.
The dynamical matter is expressed by a fermion field in the fundamental representation, and it is formulated as staggered fermions~\cite{Kogut:1974ag, Banks:1975gq} situated at lattice sites labeled by the index $x\in\{0,1,\cdots,N-1\}$, where $N =2N_P$ are the number of staggered sites associated with $N_P$ spatial lattice sites. In 1+1D, the staggered fermion formulation removes all doubles and gives only one physical fermion species.
The mass term of the staggered fermion in the Hamiltonian is given by 
\begin{equation}
    H_M \equiv \sum_{x=0}^{N-1} h_M(x) =  \sum_{x=0}^{N-1} (-1)^x\, m \,\psi^\dagger_\alpha(x) \psi_\alpha (x).
    \label{eq: HM in KS}
\end{equation}
Here, $\psi^\dagger_\alpha(x)$ ($\psi_\alpha(x)$) stands for the fermionic creation (annihilation) operator at even site $x$, while on odd sites it annihilates (creates) an anti-fermion. $m$ denotes the fermion mass\footnote{$m$ corresponds to half the fermion mass in the continuum limit.} in lattice units and $\alpha$ is the SU(2) color index that is being summed over $\alpha=1,2$.
The fermion field satisfies the anti-commutation relations $\{\psi_\alpha(x), \psi^\dagger_\beta(x')\}=\delta_{\alpha\beta}\delta_{xx'}$ and  $\{\psi_\alpha(x), \psi_\beta(x')\} = 0$.
This implies that the local fermionic Hilbert space at site $x$ is spanned by states $|f_1(x),f_2(x)\rangle$ where $f_\alpha(x)=0, 1$ is the fermion occupation number for the color index $\alpha$.
In the staggered formulation, fermion and anti-fermion excitations are given by 1 and 0 at the even and odd staggered sites of the lattice, respectively, which results in the oscillating sign for its local mass energy, $h_M(x)$. 
\begin{figure}[t!]
    \centering
    \includegraphics[scale=1]{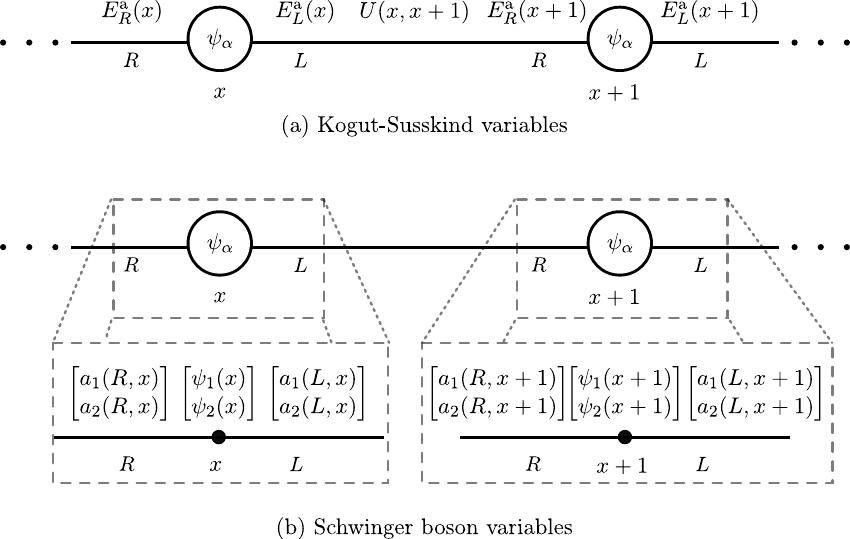}
    \caption{(a) The degrees of freedom of an SU(2) LGT in 1+1D in the KS framework are shown here for two staggered lattice sites $x$ and $x+1$. Fermionic matter in the fundamental representation residing at the lattice sites is labeled by $\psi_\alpha$, (which should not be confused with the fermion annihilation operator) with $\alpha=1,2$ being the color index. $E_L^{\rm a}$ and $E_R^{\rm a}$ are the left and the right electric field operators with adjoint color index ``${\rm a}$'' located at the corresponding links connecting sites, and they are related by the gauge link operator $U$. (b) The Schwinger boson degrees of freedom for the same SU(2) LGT are shown here overlaying the same lattice as in (a). The matter field variables, shown as SU(2) doublets at sites $x$ and $x+1$, are the same in both descriptions. However, the gauge degrees of freedom are expressed in terms of the Schwinger boson doublets $a_\alpha(s,x)$ for $s=L,R$. All degrees of freedom associated with a staggered site are shown in a box with dashed lines.}
    \label{fig: KS and Schwinger dofs}
\end{figure}
The gauge degree of freedom is located on the links connecting the staggered lattice sites, and it is formulated by working with the temporal gauge fixing condition, $A^{\rm a}_0 = 0$, where ``${\rm a}$'' is the adjoint index that takes values from $1, 2, 3$.
With this, the pair of the canonical conjugate variables associated with the spatial gauge field component are the left (right) electric field operator, $E_L^{\rm a}(x)$ $(E_R^{\rm a}(x+1))$ and the gauge holonomy or the link operator $U(x,x+1)$, which is a $2\times 2$ special unitary matrix with entries, ${U^\alpha}_{\beta}(x,x+1)$, labeled by the SU(2) color indices $\alpha$ and $\beta$.
The $U(x,x+1)$ operator resides on the link connecting the lattice sites $x$ and $x+1$, and $E_L^{\rm a}(x)$ $(E_R^{\rm a}(x+1))$ is located at the left (right) end of the same link, as shown in Fig.~\ref{fig: KS and Schwinger dofs}(a).
The left and right electric fields  are generators of gauge transformations on the link operator and related through the commutation relations
\begin{subequations}
    \begin{align} 
    \left[E^{\rm a}_L(x),{U^\alpha}_\beta(x,x+1)\right] &= -\big(t^{\rm a} U(x,x+1)\big){^\alpha}_\beta,\label{subeq: commutator EL U}\\
    \left[E^{\rm a}_R(x+1),{U^\alpha}_\beta(x,x+1)\right]&=  \big( U(x,x+1)t^{\rm a}\big){^\alpha}_\beta,
    \label{subeq: commutator ER U}
    \end{align}
    \label{eq: commutator E U}%
\end{subequations}
where $t^{\rm a} = \dfrac{\sigma^{\rm a}}{2}$, with $\sigma^{\rm a}$ being the Pauli matrices.
This constrains the electric fields residing at both ends of every link as 
\begin{equation}
    [E_L(x)]^2 = [E_R(x+1)]^2,
    \label{eq: EL ER Casimir constraint}
\end{equation}
where
\begin{equation}
    [E_s(x)]^2 = \sum_{{\rm a}=1}^3 E^{\rm a}_s(x)E^{\rm a}_s(x),
    \label{eq: Es sq Casimir definition}
\end{equation}
for $s=L,R$.
The energy of the electric field at $x$, $h_E(x)$, contributes to the total Hamiltonian as
\begin{equation}
    H_E \equiv \sum_{x=0}^{N-1} h_E(x) = \sum_{x=0}^{N-1} \frac{g^2}{4} \left\{  [E_L(x)]^2+ [E_R(x)]^2\right\},
    \label{eq: HE in KS}
\end{equation}
where $g$ is the coupling constant, and $h_E(x)$ is obtained by equating the summands\footnote{We have used a symmetric definition for the local electric energy term in the Hamiltonian using the condition in Eq.~\eqref{eq: EL ER Casimir constraint}.}.
The electric field operators satisfy the commutation relations
\begin{subequations}
    \begin{align}
        [E^{\rm a}_L(x),E^{\rm b}_{L}(x')] &= i\,\epsilon^{\rm abc}\,E^{\rm c}_L(x) \, \delta_{xx'}, \label{subeq: EL commutation algebra}\\
        [E^{\rm a}_R(x),E^{\rm b}_{R}(x')] &=  i\,\epsilon^{\rm abc}\,E^{\rm c}_R(x) \, \delta_{xx'}, \label{subeq: ER commutation algebra}
    \end{align}
  \label{eq: commutation Ea Eb}%
\end{subequations}
where $\epsilon^{\rm abc}$ are the components of the Levi-Civita tensor.
Thus, the local Hilbert space of gauge degrees of freedom at each site $x$ can be constructed from the angular momentum representation states of the SU(2) group associated with $E^{\rm a}_L(x)$ and $E^{\rm a}_R(x)$, independently, where the $[E(x)]^2$ operator in Eq.~\eqref{eq: EL ER Casimir constraint} is the Casimir operator and the azimuthal direction is taken along ${\rm a}=3$ for both $L$ and $R$ angular momentum states.
This basis, known as the electric field basis, is denoted by $\ket{j_R(x),m_R (x)}\otimes\ket{j_L(x),m_L(x)}$, where $j_L(x)$ ($j_R(x)$) and $m_L(x)$  ($m_R(x)$) are the total angular momentum quantum number and the azimuthal quantum number at the $L$ ($R$) end of the gauge link emanating from site $x$ $(x-1)$, respectively.
The quantization of angular momentum implies $j_s(x)$= 0, $1/2$, 1, $3/2$, $\cdots$, and the allowed values for the azimuthal quantum number are integers or half-integers satisfying $-j_s(x)\leq m_s(x) \leq j_s(x)$ for $s=L, R$.
However, the total angular momentum values across a link are not independent since Eq.~\eqref{eq: EL ER Casimir constraint} manifests a constraint on them:
\begin{equation}
    j_L(x)=j_R(x+1) \quad \forall x.
    \label{eq: jL jR equal constraint}
\end{equation}

Finally, the interaction energy between the matter and gauge degrees of freedom is given by the fermionic hopping term between the nearest neighbor staggered lattice sites, $x$ and $x+1$.
Color indices of the staggered fermions and the link operator $U(x,x+1)$ are contracted as
\begin{equation}
    H_I \equiv \sum_{x=0}^{N'}h_I(x,x+1) = \sum_{x=0}^{N'} \left[\psi^\dagger_\alpha(x)\, U{^\alpha}_{\beta}(x,x+1)\, \psi^\beta(x+1) + \psi^\dagger_\alpha(x+1) U{^{\dagger\alpha}}_{\beta}(x,x+1)\, \psi^\beta(x)\right],
    \label{eq: HI in KS}
\end{equation}
where $N' = N-1$ for a lattice with a periodic boundary condition (PBC), while $N' = N-2$ for an open boundary condition (OBC).
Here, $h_I(x,x+1)$ is the local interaction energy density.
The action of the $U{^\alpha}_{\beta}(x,x+1)$ operator on states in Eq.~\eqref{eq: general KS state} can be deduced from Eq.~\eqref{eq: commutator E U}\footnote{The numerical analysis performed in this work uses the LSH framework to be described in Sec.~\ref{subsubsec: LSH formulation}. Thus, for brevity, we do not provide the explicit expression for the action of $U{^\alpha}_{\beta}(x,x+1)$ operator on the basis states.
Interested readers are referred to the Refs.~\cite{Kogut:1974ag, Raychowdhury:2019iki}.
}.

The complete Hamiltonian of this theory is given by
\begin{equation}
    H = H_I + H_E + H_M,
    \label{eq: complete Hamiltonian in KS}
\end{equation}
where $H_M$, $H_E$, and $H_I$ are defined in Eqs.~\eqref{eq: HM in KS},~\eqref{eq: HE in KS}, and~\eqref{eq: HI in KS}, respectively. The complete local Hilbert space consisting of both fermion and gauge field degrees of freedom at each site $x$ is then given by $\ket{j_R,m_R}_{x} \otimes \ket{f_1, f_2}_x \otimes \ket{j_L, m_L}_x$, where the position arguments are shown as subscripts for brevity.
The full Hilbert space of the theory is spanned by states that are constructed as tensor product of the local states as
\begin{equation}
    |\Psi\rangle^{(\rm KS)} =\goldieotimes_{x=0}^{N-1} \ket{j_R,m_R}_{x} \otimes \ket{f_1, f_2}_x \otimes \ket{j_L, m_L}_x,
    \label{eq: general KS state}
\end{equation}
subjected to the constraint in Eq.~\eqref{eq: jL jR equal constraint}.
Note that, the state $\ket{j_R,m_R}_{x}$ for $x=0$ and $\ket{j_L,m_L}_{x}$ for $x=N-1$ are referred to as incoming and outgoing electric flux states, respectively. 
They play a role in characterizing the boundary conditions.
For an OBC, these states are independently fixed, while for a PBC, the incoming and outgoing electric fluxes are required to satisfy $j_R(0)=j_L(N-1)$.

The Hamiltonian matrix can be constructed in a Hilbert space spanned by the basis states given in Eq.~\eqref{eq: general KS state}.
However, a regularization is needed for PBCs for restricting the size of the Hilbert space to make its numerical computation feasible.
This is achieved by truncating the values for the total angular momentum 
at every lattice site as
\begin{equation}
    j_s(x)\leq j_{\rm max}\quad \forall x,
    \label{eq: jmax truncation in KS}
\end{equation}
for $s=L,R$.
Furthermore, $j_{\mathrm{max}}$ needs to be taken sufficiently large for studying the observables one is interested in, such that the regularization effects are negligible or systematically under control.

Not every state constructed from the basis states in Eq.~\eqref{eq: general KS state} is a physical state, since it could transform under residual time-independent gauge transformations, and thus, violate gauge invariance.
The generators of this residual symmetry at $x$ are given by the Gauss's law operators
\begin{equation}
  G^{\rm a}(x)= E^{\rm a}_L(x) + E^{\rm a}_R(x) + \psi^\dagger_\alpha(x)\left[t^{\rm a}\right]_{\alpha\beta}\psi_\beta(x).
  \label{eq: gauss law oeprator in KS}
\end{equation}
The gauge invariant sector or the physical Hilbert space is then spanned by vectors $|\Psi\rangle^{(\rm Phys)}$ that are the linear combinations of basis states in Eq.~\eqref{eq: general KS state}, subject to the Gauss's law constraint:
\begin{equation}
  G^{\rm a}(x)|\Psi\rangle^{\rm{(Phys)}}=0 \quad \forall\, {\rm a}\,,x.
  \label{eq: gauss law constraint in KS}
\end{equation}

The computational cost of numerically solving the SU(2) LGT in 1+1D in the KS framework depends on the size of the Hilbert space with the truncation in Eq.~\eqref{eq: jmax truncation in KS}, 
the construction of the physical Hilbert space using Eq.~\eqref{eq: gauss law constraint in KS}, and the action of the Hamiltonian in Eq.~\eqref{eq: complete Hamiltonian in KS} on them.
A comparative analysis of the computational cost of various formulations of this theory has been performed in Ref.~\cite{Davoudi:2020yln}.
It concluded that the LSH formulation, which uses the local gauge singlet excitations for constructing the physical Hilbert space, offers clear advantages for numerically solving the SU(2) LGT in 1+1D using classical computers\footnote{A similar conclusion for the quantum computation of the same theory is drawn in Ref.~\cite{Davoudi:2022xmb}.}.
We now briefly discuss the LSH formulation employed in this work for the numerical analysis performed in Sec.~\ref{sec: Results}.

\subsubsection{The loop-string-hadron formulation
\label{subsubsec: LSH formulation}
}

The LSH formulation uses the Schwinger boson construction of the angular momentum states~\cite{Mathur:2004kr, Mathur:2007nu, Anishetty:2009ai, Anishetty:2009nh, Mathur:2010wc} in Eq.~\eqref{eq: general KS state}.
The Schwinger bosons are described by local simple harmonic oscillator creation and annihilation operators, $a^\dagger_\alpha(s,x)$ and $a_\alpha(s,x)$, respectively, which are used for constructing the states $\ket{j_s,m_s}_x$.
The color index $\alpha=1,2$ transforms according the fundamental representation of the SU(2) gauge group, and the operators satisfy
\begin{equation}
    \left[a_\alpha (s,x), a^\dagger_\beta (s',x')\right]= \delta_{xx'}\, \delta_{ss'} \, \delta_{\alpha\beta}.
    \label{eq: schwinger boson commutation relation}
\end{equation}
The state $|j_s,m_s\rangle_x$ can then be reconstructed as
\begin{equation}
    |j_s,m_s\rangle_x = \frac{1}{\sqrt{n_1!}\sqrt{n_2!}}\left(a^\dagger_1(s,x)\right)^{n_1} \left(a^\dagger_2(s,x)\right)^{n_2}\ket{0,0}_{s,x},
    \label{eq: js ms state in schwinger bosons}
\end{equation}
where $n_1+n_2 = 2j_s$, $n_1-n_2=2m_s$, and the bare vacuum $\ket{0,0}_{s, x}$ satisfies $a_\alpha(s,x)\ket{0,0}_{s,x}=0$.
The electric field operator can be written in terms of the Schwinger bosons as
\begin{equation}
    E^{\rm a}_s(x) = a^\dagger_\alpha(s,x) \left[ t^{\rm a}\right]_{\alpha\beta}  a_\beta (s,x).
  \label{eq: EL and ER definition in Schwinger bosons}
 \end{equation}
Furthermore, the link operator $U(x,x+1)$ is split as 
\begin{equation}
    U(x,x+1) = U_L(x) U_R(x+1),
    \label{eq: U in UL and UR}
\end{equation}
such that each end of the link is re-expressed in terms of the left or right Schwinger bosons as
\begin{subequations}
    \begin{align}
         U_L(x) &= \frac{1}{\sqrt{N_L(x)+1}}
        \begin{pmatrix}
        a_2^\dagger(L,x)  & a_1 (L,x) \\
        -a_1^\dagger(L,x) & a_2(L,x)
        \end{pmatrix},
        \label{subeq: UL in schwinger bosons} \\
        U_R(x) &=
         \begin{pmatrix}
             a_1^\dagger(R,x)  &  a_2^\dagger(R,x) \\
            -a_2(R,x)  &  a_1(R,x)
         \end{pmatrix}
        \frac{1}{\sqrt{N_R(x)+1}},
        \label{subeq: UR in schwinger bosons}
   \end{align}
   \label{eq: UL and UR in schwinger bosons}%
\end{subequations}
where
\begin{equation}
    N_s(x) = a^\dagger_\alpha(s,x)  a_\alpha(s,x),
\end{equation}
denotes the number operators for $s=L,R$ oscillators\footnote{Note that, the positions of the number operators in Eq.~\eqref{eq: UL and UR in schwinger bosons} matter as the number operators do not commute with their corresponding Schwinger boson creation and annihilation operators.}.
The eigenvalues of $N_s$ are given by the number of $s-$type oscillators at $x$ and denoted as $n_s(x)$.

The Schwinger bosons are shown in Fig.~\ref{fig: KS and Schwinger dofs}(b) in comparison to the KS degrees of freedom in Fig~\ref{fig: KS and Schwinger dofs}(a).
The reconstruction of KS variables using the Schwinger bosons naturally leads to a site-local description, both for the states and the canonical variables, and allows one to identify the local gauge-invariant excitations.
This type of reformulation has been explored in a series of publications that have led to the development of the LSH framework of LGTs~\cite{Anishetty:2009ai, Anishetty:2009nh, Mathur:2004kr, Mathur:2007nu, Mathur:2010wc, Raychowdhury:2013rwa, Raychowdhury:2019iki, Kadam:2022ipf, Kadam:2024ifg}.
Its novelty lies in associating new quantum numbers to the gauge-singlet excitations constructed out of local Schwinger bosons and fermionic fields.
The Hilbert space of the theory is then spanned by the manifestly gauge-invariant states that are the tensor products of these local gauge-singlet excitations, which serve as an electric basis of the theory and belong to the physical Hilbert space, up to a remaining Abelian constraint $N_L(x) = N_R(x+1)$, to be explained below.
\begin{table}[t!]
    \renewcommand{\arraystretch}{1.85}
    \begin{center}
        \begin{tabular}{C{2.5cm}|C{2.5cm}C{4cm}C{6cm}}
            \hline
            Type & Operator label & Definition & \multicolumn{1}{c} {LSH representation}\\
            \hline
            \hline
            \multirow{4}{*} {\rotatebox{90}{\parbox{2.5cm}{\centering Pure gauge loop operators}}} & $\mathcal{L}^{++}$ & $a^\dagger_{\alpha}(R)\, a^\dagger_\beta(L)\, \epsilon_{\alpha\beta}$ & $\Lambda^+ \sqrt{(N_l + 1) (N_l + 2 + (N_i \oplus N_o))}$\\
            & $\mathcal{L}^{--}$ & $a_{\alpha}(R)\, a_\beta(L)\, \epsilon_{\alpha\beta}$ &  $\Lambda^- \sqrt{N_l (N_l + 1 + (N_i \oplus N_o))}$\\
            & $\mathcal{L}^{+-}$ & $a^\dagger_{\alpha}(R)\, a_\beta(L)\, \delta_{\alpha\beta}$ & $- \chi_i^\dagger \, \chi_o$\\
            & $\mathcal{L}^{-+}$ & $a_{\alpha}(R)\, a^\dagger_\beta(L)\, \delta_{\alpha\beta}$ & $\chi_i \, \chi_o^\dagger$\\
            \hline
            \multirow{4}{*} {\rotatebox{90}{\parbox{3cm}{\centering Incoming string operators}}} & $\mathcal{S}^{++}_{\rm in}$ & $a^\dagger_{\alpha}(R)\, \psi^\dagger_\beta\, \epsilon_{\alpha\beta}$ & $\chi_i^\dagger \, (\Lambda^+)^{N_o}\,\sqrt{N_l + 2 - N_o}$\\
            &  $\mathcal{S}^{--}_{\rm in}$ & $a_{\alpha}(R)\, \psi_\beta\, \epsilon_{\alpha\beta}$ &  $\chi_i \, (\Lambda^-)^{N_o} \, \sqrt{N_l + 2(1 - N_o)}$\\
            &  $\mathcal{S}^{+-}_{\rm in}$ & $a^\dagger_{\alpha}(R)\, \psi_\beta\, \delta_{\alpha\beta}$ & $ \chi_o  \, (\Lambda^+)^{1-N_i}\sqrt{N_l+1+N_i} $\\
            &  $\mathcal{S}^{-+}_{\rm in}$ & $a_{\alpha}(R)\, \psi^\dagger_\beta\, \delta_{\alpha\beta}$ & $\chi_o^\dagger \, (\Lambda^-)^{1-N_i} \sqrt{N_l+2 N_i}$\\
            \hline
            \multirow{4}{*} {\rotatebox{90}{\parbox{3cm}{\centering Outgoing string operators}}} & $\mathcal{S}^{++}_{\rm out}$ & $\psi^\dagger_{\alpha}\, a^\dagger_\beta(L)\, \epsilon_{\alpha\beta}$ & $\chi_o^\dagger \, (\Lambda^+)^{N_i} \,\sqrt{N_l + 2 - N_i}$\\
            & $\mathcal{S}^{--}_{\rm out}$ & $\psi_{\alpha}\, a_\beta(L)\, \epsilon_{\alpha\beta}$ & $\chi_o \, (\Lambda^-)^{N_i} \,\sqrt{N_l + 2(1 - N_i)}$\\
            & $\mathcal{S}^{+-}_{\rm out}$ & $\psi^\dagger_{\alpha}\, a_\beta(L)\, \delta_{\alpha\beta}$ & $\chi_i^\dagger \, (\Lambda^-)^{1-N_o} \sqrt{N_l+2 N_o}$\\
            & $\mathcal{S}^{-+}_{\rm out}$ & $\psi_{\alpha}\, a^\dagger_\beta(L)\, \delta_{\alpha\beta}$ & $\chi_i \, (\Lambda^+)^{1-N_o} \sqrt{N_l+1+N_o}$\\
            \hline
            \rule{0pt}{0.6cm}\multirow{2}{*} {\rotatebox{90}{\parbox{1.8cm}{\centering Hadron operators}}} & $\mathcal{H}^{++}$ & $-\frac{1}{2}\,\psi^\dagger_{\alpha}\, \psi^\dagger_\beta\, \epsilon_{\alpha\beta}$ & $\chi_i^\dagger\,\chi_o^\dagger$\\
            \rule{0pt}{0.7cm}& $\mathcal{H}^{--}$ & $\frac{1}{2}\,\psi_{\alpha}\, \psi_\beta\, \epsilon_{\alpha\beta}$ & $-\chi_i\,\chi_o$\\
            \hline
            \multirow{3}{*} {\rotatebox{90}{\parbox{2cm}{\centering Number operators}}} & $N_L$ & $a^\dagger_{\alpha}(L)\,a_\beta(L) \, \delta_{\alpha\beta}$ & $N_l+N_o(1-N_i)$\\
            & $N_R$ & $a^\dagger_{\alpha}(R)\,a_\beta(R) \, \delta_{\alpha\beta}$ & $N_l+N_i(1-N_o)$\\
            & $N_\psi$ & $\psi^\dagger_{\alpha}\,\psi_\beta \, \delta_{\alpha\beta}$ & $N_i+N_o$\\
            \hline
            \hline           
        \end{tabular}
    \end{center}
    \caption{List of the local gauge-invariant operators and their corresponding LSH operators. All the position labels are omitted here for brevity. The first column denotes the type of operator, loop-type, string-type, hadron-type or number operators. The second column indicates the operator label, and the third column provides their definitions. The subscripts `in' and `out' in labeling the string operators stand for the incoming and outgoing type of electric fluxes from the $R$ and $L$ ends of the link associated with site $x$, respectively. Finally, the last column expresses the operators in terms of creation, annihilation, and number operators for the LSH quantum numbers that are defined in Eqs.~\eqref{eq: Lambda raising lowering definition},~\eqref{eq: LSH fermion creation and annihilation operators}, and~\eqref{eq: LSH number operator definitions}, respectively. Here, $(\Lambda^\pm)^{N_q} \equiv (1-N_{q}) + \Lambda^\pm N_{q}$ and $(\Lambda^\pm)^{1-N_q} \equiv \Lambda^\pm (1-N_{q}) +  N_{q}$ denote the conditional operators for $q=i,o$.
    \label{tab: LSH operators}}
\end{table}
A complete list of gauge-singlet operators is provided and shown to form a closed algebra in Ref.~\cite{Raychowdhury:2019iki}, and we summarize those operators in Table~\ref{tab: LSH operators}.
The local gauge-invariant Hilbert space is spanned by states with the quantum numbers corresponding to the excitation numbers of only creation-type operators: $\mathcal{L}^{++}(x)$, $\mathcal{S}^{++}_{\rm in}(x)$, $\mathcal{S}^{++}_{\rm out}(x)$, and $\mathcal{H}^{++}(x)$.
The first operator $\mathcal{L}^{++}(x)$ contains only the Schwinger boson operators, and thus, obeys boson statistics for its occupation number $n_l(x)$ that takes non-negative integer values $n_l(x)\in \{0,1,2,\cdots\}$.
This operator is called the loop operator.
The next two operators $\mathcal{S}^{++}_{\rm in}(x)$ and $\mathcal{S}^{++}_{\rm out}(x)$ are mixed operators containing a fermion operator and a Schwinger boson operator, and hence, referred to as string operators, while the last operator $\mathcal{H}^{++}(x)$ is composed entirely of fermionic operators and known as the hadron operator.
The presence of a fermion operator in the string and hadron excitations causes the quantum numbers associated with them, $n_i$ and $n_o$, to respect the fermion statistics, i.e. $n_i, n_o \in \{0,1\}$.
The basis of the local Hilbert space is spanned by the states labeled with the LSH quantum numbers $n_l$, $n_i$, and $n_o$, that are defined as
\begin{subequations}
  \begin{align}
    \ket{n_l, n_i=0, n_o=0}_x &\equiv \mathcal{N}_{n_l, 0, 0} \; (\mathcal{L}^{++}(x))^{n_l} \ket{0}_x, \label{subeq: local LSH state nl 0 0 definition}\\
    \ket{n_l, n_i=0, n_o=1}_x &\equiv \mathcal{N}_{n_l, 0, 1} \; (\mathcal{L}^{++}(x))^{n_l} \mathcal{S}^{++}_{\rm out}(x) \ket{0}_x,\label{subeq: local LSH state nl 0 1 definition}\\
    \ket{n_l, n_i=1, n_o=0}_x &\equiv \mathcal{N}_{n_l, 1, 0} \; (\mathcal{L}^{++}(x))^{n_l} \mathcal{S}^{++}_{\rm in}(x) \ket{0}_x, \label{subeq: local LSH state nl 1 0 definition}\\
    \ket{n_l, n_i=1, n_o=1}_x &\equiv \mathcal{N}_{n_l, 1, 1} \;  (\mathcal{L}^{++}(x))^{n_l} \mathcal{H}^{++}(x)\ket{0}_x , \label{subeq: local LSH state nl 1 1 definition}
  \end{align}
  \label{eq: local LSH state definitions}%
\end{subequations}
where $\ket{0}_x$ is the local projection of the normalized unexcited state at position $x$, i.e. $\prod_x\braket{0|0}_x=1$, which is annihilated by all $a_\alpha(s,x)$'s and $\psi_\alpha(x)$'s.
The states are normalized to one with the normalization constant $\mathcal{N}_{n_l,n_i,n_o}$ given by 
\begin{equation}
    \mathcal{N}_{n_l,n_i,n_o} \equiv \frac{1}{\sqrt{n_l! \ (n_l+1+(n_i \oplus n_o ))!}}, \label{eq: normalization definition}
\end{equation}
where $\oplus$ denotes addition modulo 2.

The action of the operators in Table~\ref{tab: LSH operators} can be expressed in terms of creation and annihilation operators associated with the LSH quantum numbers, as shown in the third column.
Here, $\Lambda^{+}$ and $\Lambda^{-}$ are the bosonic ladder operators associated with the quantum number $n_l$:
\begin{equation}
  _{x'}\bra{n_l^\prime , n_i^\prime, n_o^\prime } \Lambda^{\pm}(x) \ket{n_l,n_i,n_o}_x = \delta_{n_l^\prime, n_l \pm 1} \delta_{n_i^\prime n_i} \delta_{n_o^\prime n_o}\delta_{x x'}.
  \label{eq: Lambda raising lowering definition}
\end{equation}
The fermionic raising (lowering) operators $\chi^\dagger_q(x)$ ($\chi_q(x)$) with $q=i$ and $q=o$ labeling the $n_i$ and $n_o$ quantum numbers, respectively.
These gauge-invariant operators obey the fermion anti-commutation relations $\{ \chi_{q'}(x') ,\chi_{q}(x) \} = 0$ and $\{ \chi_{q'}(x'), \chi_{q}^\dagger(x)\}  = \delta_{qq'}\delta_{xx'}$, and thus, 
\begin{subequations}
    \begin{align}
        _{x'}\bra{n_l^\prime , n_i^\prime, n_o^\prime } \chi^\dagger_i(x) \ket{n_l,n_i,n_o}_x &= \delta_{n_l^\prime n_l} \delta_{n_i^\prime ,n_i+1} \delta_{n_o^\prime n_o}\delta_{xx'} , \label{subeq: ni creation dagger definition}\\
        _{x'}\bra{n_l^\prime , n_i^\prime, n_o^\prime } \chi_i(x) \ket{n_l,n_i,n_o}_x &= \delta_{n_l^\prime n_l} \delta_{n_i^\prime ,n_i-1} \delta_{n_o^\prime n_o}\delta_{x x'} , \label{subeq: ni annihilition dagger definition}\\
        _{x'}\bra{n_l^\prime , n_i^\prime, n_o^\prime } \chi^\dagger_o(x) \ket{n_l,n_i,n_o}_x &= \delta_{n_l^\prime n_l} \delta_{n_i^\prime n_i} \delta_{n_o^\prime ,n_o+1}\delta_{xx'} , \label{subeq: no creation dagger definition}\\
        _{x'}\bra{n_l^\prime , n_i^\prime, n_o^\prime } \chi_o(x) \ket{n_l,n_i,n_o}_x &= \delta_{n_l^\prime n_l} \delta_{n_i^\prime n_i} \delta_{n_o^\prime ,n_o-1}\delta_{xx'} . \label{subeq: no annihilition dagger definitions}
    \end{align}
    \label{eq: LSH fermion creation and annihilation operators}%
\end{subequations}
The number operators corresponding to the LSH quantum numbers are defined as
\begin{subequations}
    \begin{align}
        N_l(x) &\equiv \Lambda^+(x)\Lambda^-(x) , \label{subeq: Nl definition}\\
        N_i(x) &\equiv \chi_i^\dagger(x)\chi_i(x) , \label{subeq: Ni definition}\\
        N_o(x) &\equiv \chi_o^\dagger(x)\chi_o(x) , \label{subeq: No definition}
    \end{align}
    \label{eq: LSH number operator definitions}%
\end{subequations}
The Hilbert space of the theory is constructed from the states that are tensor products of the local LSH states defined in Eq.~\eqref{eq: local LSH state definitions}.
Although, such states satisfy the Gauss's law condition in Eq.~\eqref{eq: gauss law constraint in KS} by virtue of their construction from gauge-invariant operators, they do not manifestly satisfy the constraint in Eq.~\eqref{eq: EL ER Casimir constraint} which relates the $L$ and $R$ electric fields across a link.
Thus, the physical states in the LSH framework need to have this constraint imposed, which manifests as an constraint on the operators $N_L$ and $N_R$ at neighboring sites as
\begin{equation}
    N_L(x) = N_R(x+1) \quad \forall x,
    \label{eq: Abelian gauss law constraint on NL NR}
\end{equation}
or equivalently on their eigenvalues as
\begin{equation}
    n_L(x) = n_l(x) + n_o(x)(1-n_i(x)) = n_l(x+1) + n_i(x+1)(1-n_o(x+1)) = n_R(x+1) \quad \forall x.
    \label{eq: Abelian gauss law constraint on LSH qunatum numbers}
\end{equation}
Here, the relation between $n_s(x)$ and the LSH quantum numbers is obtained from the number operators shown in Table~\ref{tab: LSH operators}.
The eigenvalue of the $N_s$ number operator is related to the corresponding total angular momentum quantum number in Eq.~\eqref{eq: general KS state} as
\begin{equation}
    n_s(x) = 2 j_s(x).
    \label{eq: NL NR in terms of jL jR}
\end{equation}
This implies that the definitions for the PBCs and OBCs are also naturally carried over to the LSH quantum numbers $n_l$, $n_i$, and $n_o$ at the first and the last lattice sites.

Equations~\eqref{eq: Abelian gauss law constraint on NL NR} and~\eqref{eq: Abelian gauss law constraint on LSH qunatum numbers} are called the Abelian Gauss's law conditions, as they are algebraic conditions that can be simultaneously imposed on the states.
They are obtained by substituting Eq.~\eqref{eq: EL and ER definition in Schwinger bosons} into Eq.~\eqref{eq: EL ER Casimir constraint} and using the operator definitions in Table~\ref{tab: LSH operators}.
Thus, the physical Hilbert space of the theory is spanned by the states
\begin{equation}
    |\Psi\rangle^{(\rm Phys)} = \goldieotimes_{x=0}^{N-1} \ket{n_l,n_i,n_o}_x,
    \label{eq: general LSH state}
\end{equation}
that are subjected to the constraint in Eq.~\eqref{eq: Abelian gauss law constraint on LSH qunatum numbers}.
Furthermore, the truncation condition in Eq.~\eqref{eq: jmax truncation in KS} can be imposed on the LSH Hilbert space as
\begin{equation}
    n_s(x)\leq 2 j_{\rm max} \quad \forall x,
    \label{eq: jmax truncation on Ns}
\end{equation}
for $s=L,R$, and the dimension of the regularized physical Hilbert space remains the same in both LSH and KS formulations.

Finally, the Hamiltonian can be expressed in terms of the LSH operators as shown below.
Utilizing Table~\ref{tab: LSH operators} and Eq.~\eqref{eq: NL NR in terms of jL jR} in Eqs.~\eqref{eq: HM in KS} and~\eqref{eq: HE in KS}, respectively, one finds that the mass energy and the electric energy at site $x$ is given by
\begin{equation}
    h_M(x) = (-1)^x \, m \,N_\psi, \label{eq: hM in LSH}
\end{equation}
and
\begin{equation}
    h_E(x) = \frac{g^2}{4} \left[ \frac{1}{2} N_R(x) \left(\frac{1}{2} N_R(x) + 1 \right) + \frac{1}{2} N_L(x) \left(\frac{1}{2} N_L(x) + 1 \right) \right] , \label{eq: hE in LSH}
\end{equation}
respectively.
The local interaction energy contribution, $h_I(x,x+1)$, is given by
\begin{align}
    h_I(x,x+1) &= \frac{1}{\sqrt{N_L(x)+1}} \left[ \mathcal{S}_{\rm out}^{++}(x)  \mathcal{S}_{\rm in}^{+-}(x+1) + \mathcal{S}_{\rm out}^{+-}(x)\mathcal{S}_{\rm in}^{--}(x+1)  \right] \frac{1}{\sqrt{N_R(x+1)+1}}\nonumber\\
    &\quad + \frac{1}{\sqrt{N_R(x+1)+1}} \left[ \mathcal{S}_{\rm in}^{-+}(x+1) \mathcal{S}_{\rm out}^{--}(x)+ \mathcal{S}_{\rm in}^{++}(x+1) \mathcal{S}_{\rm out}^{-+}(x)  \right] \frac{1}{\sqrt{N_L(x)+1}},
    \label{eq: hI in LSH}
\end{align}
where the terms in the second line are Hermitian conjugates of the terms in the first line.  
Equation~\eqref{eq: hI in LSH} is obtained by substituting Eqs.~\eqref{eq: U in UL and UR} and~\eqref{eq: UL and UR in schwinger bosons} into Eq.~\eqref{eq: HI in KS}, and using the operator definitions provided in Table~\ref{tab: LSH operators}.
Note that, the square root factors that include the $N_L(x)$ and $N_R(x+1)$ terms are equal to each other as the Hamiltonian matrix is constructed only in the physical Hilbert space where the states obey the Abelian Gauss's law constraint in Eq.~\eqref{eq: Abelian gauss law constraint on NL NR}.

Equations~\eqref{eq: hM in LSH}-\eqref{eq: hI in LSH} express the Hamiltonian in Eq.~\eqref{eq: complete Hamiltonian in KS} in terms of the LSH operators defined in Eqs.~\eqref{eq: Lambda raising lowering definition}-\eqref{eq: LSH number operator definitions} and listed in Table~\ref{tab: LSH operators}.
In the remaining part of this section, we explain the discrete symmetries of this Hamiltonian, which are used in performing the numerical analysis in Sec.~\ref{sec: Results}.
\subsubsection{Global symmetries of an SU(2) lattice gauge theory Hamiltonian in 1+1D
\label{subsubsec: symmetries}
}
In this work, we restrict ourselves to studying the SU(2) LGT with a non-zero fermion mass, $m\neq0$, as the $m=0$ case is an integrable theory that is exactly solvable~\cite{Witten:1983ar,Huang:2021yjh}.
The Hamiltonian in Eq.~\eqref{eq: complete Hamiltonian in KS} commutes with two global charges: the total baryon number, $\mathcal{B}$, and  the lattice electric charge, $\Delta \mathcal{Q}$.
The former is defined as the difference between the total numbers of particle and anti-particle excitations, and for a staggered fermion lattice with SU(2) color index, it is given by
\begin{equation}
    \mathcal{B} \equiv \left(\sum_{x=0}^{N-1} N_\psi(x)\right) -N = \left(\sum_{x=0}^{N-1} \left[N_i(x) + N_o(x)\right]\right) - N.
    \label{eq: baryon number B definition}
\end{equation}
The latter is defined as the total electric flux sourced or sank by the lattice, which is expressed as 
\begin{equation}
    \Delta \mathcal{Q} \equiv  N_L(N-1) - N_R(0).
    \label{eq: Delta Q sq definition}
\end{equation}
For OBCs, the lattice charge can take integer values, $\Delta \mathcal{Q} \in \{-N,-N+1,\cdots,N\}$ as the fermion occupation at each site can only source or sink a half-integer (one unit) of electric field flux $j_s(x)$ ($n_s(x)$).
We also note that, $\Delta \mathcal{Q}$ can take even (odd) values if and only if $\mathcal{B}$ is even (odd).
Furthermore, it is clear from the discussion below Eq.~\eqref{eq: general KS state} that the value of $\Delta \mathcal{Q}$ must be zero for PBCs, and thus, only $\mathcal{B}$ is the relevant conserved charge associated with a global symmetry for lattices with PBCs.

Next, we look at the symmetries of the Hamiltonian that are associated with the transformations of the staggered spatial lattice.
The Hamiltonian with a PBC is invariant under the spatial translation of the lattice when the staggered lattice is translated by an even distance in units of the lattice spacing.
The physical momenta of the lattice, $k\in \frac{\pi}{N} \big\{-\frac{N}{2},-\frac{N}{2}+1, \cdots, \frac{N}{2}-1 \big\}$, are given by the eigenvalues of the momentum matrix that generates these translations.
The translation operation considered in this work for constructing the eigenstates of the momentum operator is given by translating the lattice in the forward direction by two units, and it is denoted by an operator $\mathcal{T}_2$.
The LSH states transform under ${\mathcal{T}}_2$ as
\begin{equation}
    {\mathcal{T}}_2 \, |\Psi\rangle^{(\rm Phys)} = \goldieotimes_{x=0}^{N-1} {\mathcal{T}}_2 \ket{n_l,n_i,n_o}_x = (-)^{\mathcal{T}_2}\,\goldieotimes_{x=0}^{N-1} \ket{n'_l,n'_i,n'_o}_{x},
    \label{eq: translation on LSH state}
\end{equation}
where $n'_{l/i/o}(x) = n_{l/i/o}(x-2)$ in the last term denotes that the values of LSH quantum numbers at $x$ are translated to the state at site $x+2$ modulo $N$.
Furthermore, $(-)^{\mathcal{T}_2}$ denotes the phase from rearranging the translated fermion creation operators to restore the chosen convention of fermion operator ordering.
For example, if the convention is chosen to order the fermionic creation operators for the corresponding LSH quantum numbers to be in ascending order of the lattice site index and $\chi^\dagger_i$ before $\chi^\dagger_o$ from left to right, then the translation of fermion creation operators at the last two sites, if present, would require reordering that will lead to an overall $+1$ or $-1$ phase depending on the number of fermion creation operator exchanges.   

The lattice translation operator maps to the continuous spatial translation symmetry in the continuum limit.
Apart from that, the spatial parity and the charge conjugation are the discrete symmetries of a continuum gauge theory in 1+1D, and they are also realized on the staggered lattice as discussed below.
The lattice parity transformation, ${\mathcal{P}}$, is defined as a reflection about an axis that passes through one of the lattice sites.
This parity transformation itself is a symmetry of the Hamiltonian in Eq.~\eqref{eq: complete Hamiltonian in KS} only for PBCs, and thus, different axis choices are equivalent since they are related through the translational symmetry.
Here, we choose the axis going through the lattice site at $N/2-1$, and let $x=N/2-1+i$ that goes to $x'=N/2-1-i = x-2i$ after the reflection.
Then the parity transformation on the LSH states is given by
\begin{equation}
    {\mathcal{P}} \, |\Psi\rangle^{(\rm Phys)} = \goldieotimes_{x=0}^{N-1} {\mathcal{P}}\ket{n_l,n_i,n_o}_x = (-)^\mathcal{P}\,\goldieotimes_{x=0}^{N-1} \; \eta^{\mathcal{P}}_{x} \ket{n'_l,n'_i,n'_o}_{x},
    \label{eq: parity on LSH state}
\end{equation}
where $(-)^\mathcal{P}$ is the fermion re-ordering phase similar to $(-)^{\mathcal{T}_2}$ and $\eta^{\mathcal{P}}_{x} = (-1)^{n'_l(x)+(x+1)(n'_o(x)+n'_i(x))}$.
The transformed fermionic LSH quantum numbers are given by $n'_l(x) = n_l(x)$, $n'_i(x) = n_o(x-2i)$, and $n'_o(x) = n_i(x-2i)$.
Finally, the lattice charge conjugation operator, ${\mathcal{C}}$, acts on the LSH states as
\begin{equation}
    {\mathcal{C}} \, |\Psi\rangle^{(\rm Phys)} = \goldieotimes_{x=0}^{N-1} {\mathcal{C}}\ket{n_l,n_i,n_o}_x = (-)^\mathcal{C}\,\goldieotimes_{x=0}^{N-1} \; \eta^{\mathcal{C}}_{x} \ket{n'_l,n'_i,n'_o}_{x}.
    \label{eq: CC on LSH state}
\end{equation}
where $(-)^\mathcal{C}$ is again the fermion re-ordering phase.
Furthermore, $n'_l(x)=n_l(x-1)$, $n'_i(x) = 1 \oplus n_o(x-1)$ and $n'_o(x) = 1 \oplus n_i(x-1)$, and $\eta^{\mathcal{C}}_x=1$ if $n'_i(x)=n'_o(x)$, otherwise, $\eta^{\mathcal{C}}_x=(-1)^{x+n'_o}$.
An example of spatial translation, parity and charge conjugation operators, ${\mathcal{T}}_2$, ${\mathcal{P}}$, and ${\mathcal{C}}$, respectively, acting on an LSH state is shown here, assuming a periodic lattice.
\begin{subequations}
\begin{align}
    \ket{\Psi} & =  \ket{0,0,1}_0\otimes\ket{1,0,0}_1\otimes\ket{1,1,1}_2\otimes\ket{0,1,0}_3 , \label{eq: sample state}\\
    {\mathcal{T}}_2\ket{\Psi} &= (-1) \ket{1,1,1}_0\otimes\ket{0,1,0}_1 \otimes \ket{0,0,1}_2\otimes\ket{1,0,0}_3 ,\label{eq: sample state Translation}\\
    {\mathcal{P}}\ket{\Psi} &= (+1) \ket{1,1,1}_0\otimes\ket{1,0,0}_1 \otimes \ket{0,1,0}_2\otimes\ket{0,0,1}_3 ,\label{eq: sample state Parity}\\
    {\mathcal{C}}\ket{\Psi} &= (-1) \ket{0,1,0}_0\otimes \ket{0,0,1}_1\otimes\ket{1,1,1}_2 \otimes \ket{1,0,0}_3.\label{eq: sample state Charge Conjugation}%
\end{align}
\end{subequations}
Equations~\eqref{eq: sample state Translation},~\eqref{eq: sample state Parity}, and \eqref{eq: sample state Charge Conjugation} are obtained by performing the transformations in Eqs.~\eqref{eq: translation on LSH state},~\eqref{eq: parity on LSH state}, and~\eqref{eq: CC on LSH state}, on the sample state $\ket{\Psi}$ given in Eq.~\eqref{eq: sample state}, respectively.
The transformations in Eqs.~\eqref{eq: translation on LSH state}-\eqref{eq: CC on LSH state} have been derived using their corresponding transformations on the field operators, as discussed in Appendix~\ref{app: symmetry on operators}.
Moreover, these transformations have been numerically verified to be the symmetries of the Hamiltonian in Eq.~\eqref{eq: complete Hamiltonian in KS} in a truncated Hilbert space for various values of $j_{\rm max}$.
The ${\mathcal{T}}_2$, ${\mathcal{P}}$, and ${\mathcal{C}}$ operations have been utilized in Sec.~\ref{sec: Results} for studying the quantum chaos in an SU(2) LGT in 1+1D with fermionic matter.

\section{Results of the ETH tests
\label{sec: Results}
}
In this section, we present results for various ETH markers obtained from using the LSH formulation explained in the previous section.
We first discuss indicators of quantum chaos and identify the Hamiltonian parameter space where quantum chaos is manifest for system sizes that are numerically accessible for us.
We then calculate the variances in both the diagonal and off-diagonal MEs for local and extended operators, and show the exponential scaling with the system size.
Furthermore, we study at how late time the MEs relative to the microcanonical values show complete GOE behavior.
Finally, we obtain the $f_O$ functions for both local and extended operators.
\subsection{Indicators of quantum chaos
\label{subsec: quantum chaos}
}
\subsubsection{Level statistics
\label{subsubsec: level statistics}
}
The distribution of energy level spacings indicates whether a quantum system is integrable or exhibits quantum chaos~\cite{Bohigas:1983er}.
A Wigner-Dyson statistics for level spacings is considered as a defining property of a quantum chaotic system, while a Poisson statistics indicates an integrable quantum system.
The Wigner-Dyson distribution does not have a closed analytical form.
However, it closely agrees with the distribution of level spacings in a random $2\times2$ matrix ensemble. 
For a chaotic system described by a Hermitian time-reversal-invariant Hamiltonian, the corresponding random matrix ensemble is the Gaussian orthogonal ensemble (GOE). 
However, to use the distribution of level spacings as an indicator of quantum chaos, one has to lift all discrete symmetries in the Hamiltonian. 
The total Hamiltonian of a system with discrete symmetries can be block diagonalized, where each block is uniquely labeled by the quantum numbers corresponding to those symmetries.
The different blocks are uncorrelated, so mixing the eigenvalues from different blocks conceal the quantum chaotic nature of the Hamiltonian. The mixed eigenvalues often exhibit a Poisson statistics for level spacing which is a characteristic of an integrable system~\cite{Santos:2010qxi,DAlessio:2015qtq}.  

Throughout Sec.~\ref{sec: Results}, we work with a PBC and choose the Hamiltonian block that is translationally invariant (zero momentum) and is also even under both parity and charge conjugation. 
Furthermore, we fix the total baryon number to $\mathcal{B}=0$. Because of this, we demand fewer computational resources than performing the same calculation in the full Hilbert space, which allows us to study lattice sizes of up to $N=14$ with the computational resources available to us.
The energy eigenvalues of this subspace Hamiltonian are arranged in ascending order and they are shifted by a constant energy such that the ground state energy is zero.
We denote these eigenvalues by $E_a$, where $a\in\{0,1,\cdots, \mathcal{D}-1\}$ with $\mathcal{D}$ being the dimension of the chosen symmetry sector, and $E_0=0$.
Furthermore, for a finite size system, the eigenstates near the edges of the spectrum are not thermal, and thus, they are omitted from the following analysis~\cite{DAlessio:2015qtq}.
We do so by using eigenenergies $E_a$ that fall in the range
\begin{align}
    \frac{ E_0 - E_{\rm avg}}{2}\,<\,E_a-E_{\rm avg} \,<\,  \frac{E_{\mathcal{D}-1} - E_{\rm avg}}{2},
    \label{eq: energy window def}
\end{align}
where $ E_{\rm avg}$ is the mean energy.

\begin{figure}[t]
\centering
\includegraphics[scale=1.0]{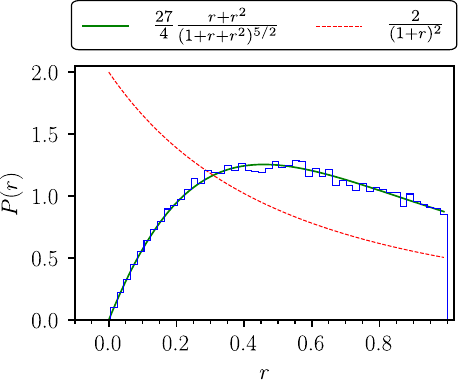}
\caption{Shown here is a histogram (blue) of the probability density of restricted gap ratios compared to the GOE (green solid) and the Poisson (red dashed) predictions. The histogram is numerically obtained for the Hamiltonian in Eq.~\eqref{eq: complete Hamiltonian in KS} in LSH formulation for $N=14$ with the parameters $g^2=0.25$ and $m=0.25$, and local Hilbert space truncation  $j_{\mathrm{max}}=1/2$. 
\label{fig: restricted gap ratio}}
\end{figure}

\begin{figure}[tbp] 
\centering
\includegraphics[scale=1.0]{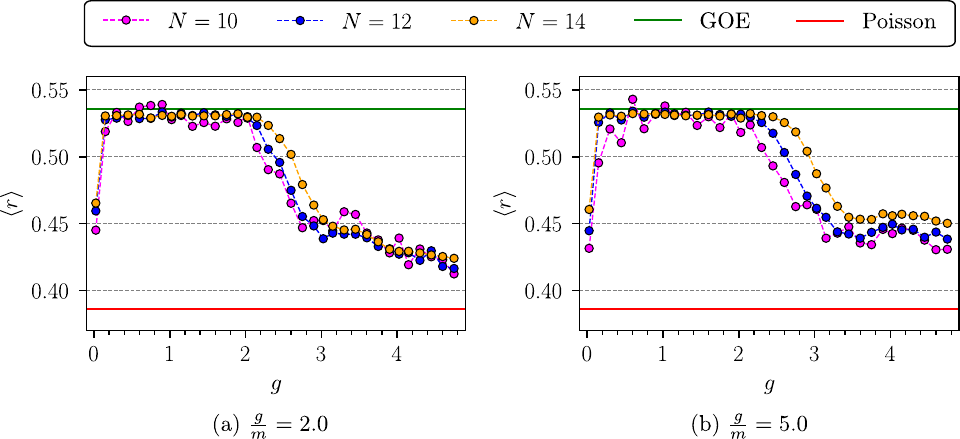}
\caption{The mean restricted gap ratio, $\langle r \rangle$, is shown for various couplings $g$ and the system sizes $N=10$, $12$ and $14$, where $m=g/2$ in (a) and $m=g/5$ in (b). For all data points in this figure, the local Hilbert space is truncated with $j_{\mathrm{max}}=1/2$.
The green (red) line denotes the value predicted by the GOE (Poisson ensemble), which indicates that the system is chaotic (integrable).
We note that for non-zero, finite values of $g$ and $m$, the LGT considered here remains chaotic for a range of parameters. We utilize this in Sec~\ref{sec: Results} to ensure that the parameters are chosen from the chaotic region.
\label{fig: gap ratio param scan}}
\end{figure}

One of the indicators of quantum chaos is the restricted ratio of level spacing, $r_a$, given by
\begin{equation}
    0 \leq r_a = \frac{ \min [ \delta_a, \delta_{a-1} ] }{ \max  [ \delta_a, \delta_{a-1} ] } \leq 1,
    \label{eq: restricted gap ratio defn}
\end{equation}
where $\delta_a = E_{a+1}-E_a$.
The prediction for the distribution of the restricted gap ratio from averaging over a GOE of $3\times3$ matrices is given by~\cite{Oganesyan_2007, Schonle:2020grk}
\begin{equation}
    P(r) = \frac{27}{4} \frac{ r + r^2}{ ( 1 + r + r^2 )^{5/2}},
    \label{eq: restricted gap ratio GOE distribution}
\end{equation}
where $r$ is a continuous variable.
A comparison between Eq.~\eqref{eq: restricted gap ratio GOE distribution} and our results for the Hamiltonian with parameters $m=0.25$ and $g^2=0.25$ and the local Hilbert space truncation $j_{\rm max} = 1/2$ on a $N=14$ lattice size
can be found in Fig.~\ref{fig: restricted gap ratio}. 
The agreement is good, up to small statistical fluctuations, which indicates that the system is quantum chaotic at these parameters.\footnote{One can also use the distribution of level spacings as an indicator and compare with the GOE prediction. In this case, one needs to unfold the spectrum, since the mean level spacing of the GOE is one, while that of a chaotic Hamiltonian depends on the energy and Hamiltonian parameters.
We explain an unfolding procedure and show the agreement between the distribution of the level spacings and the GOE prediction in Appendix~\ref{app: level statistics}.}
We have also plotted the distribution obtained from the Poisson statistics of an integrable theory, $P(r) = \frac{2}{(1+r)^2}$, as a dotted red line for contrast.
The expectation value of the restricted gap ratio over the distribution, $\braket{r}$, can also be used as an indicator of quantum ergodicity.
Its value for the distribution in Eq.~\eqref{eq: restricted gap ratio GOE distribution} is $\braket{r}=0.5359$, while for an infinitely constrained integrable system it is $\braket{r}=0.3863$~\cite{Atas_2013}.
Furthermore, to depict the range of Hamiltonian parameters over which the theory remains quantum chaotic for a given lattice size, we plot the value of $\braket{r}$ against the coupling constant $g$ in Fig.~\ref{fig: gap ratio param scan}, for two different ratios of $g/m$ with the local Hilbert space restricted to $\jmax = 1/2$.
First, we note that at a given $N$ and a given ratio $g/m$, the theory approaches integrability for large values of $g$, and thus in turn, large values of $m$.
This behavior is expected since in these parameter regions the dominant contribution to the Hamiltonian comes from the diagonal part, and the non-diagonal mixing of the Hilbert space is relatively suppressed.
Moreover, when the coupling constant $g$ and mass $m$ are nearly zero, the theory approaches an integrable limit since it is perturbatively solvable using bosonization~\cite{Witten:1983ar,Huang:2021yjh}.
However, at a fixed finite coupling, enlarging the lattice size enhances the non-integrability of the theory. We expect that at any nonzero and finite values of $g$ and $m$, the theory is non-integrable in the thermodynamic limit (infinite volume).

So far, we considered the crudest truncation of the local Hilbert space with $\jmax=1/2$.
With increasing values of $\jmax$, the transitions among the newly added states in the Hilbert space experience a relative suppression due to the large electric energy gap compared with the fermion hopping energy.
On the other hand, the spectrum up to a fixed energy for a given lattice converges when $\jmax$ reaches a critical value, and thus, its quantum ergodic characteristics in the same low energy region are independent of any further increase in $\jmax$.
The $\jmax$ dependence of the spectrum can be found in Appendix~\ref{app: spectrum jmax}.

\subsubsection{Spectral form factors
\label{subsubsec: SFF}
}

\begin{figure}[t!]
\includegraphics[scale=1.0]{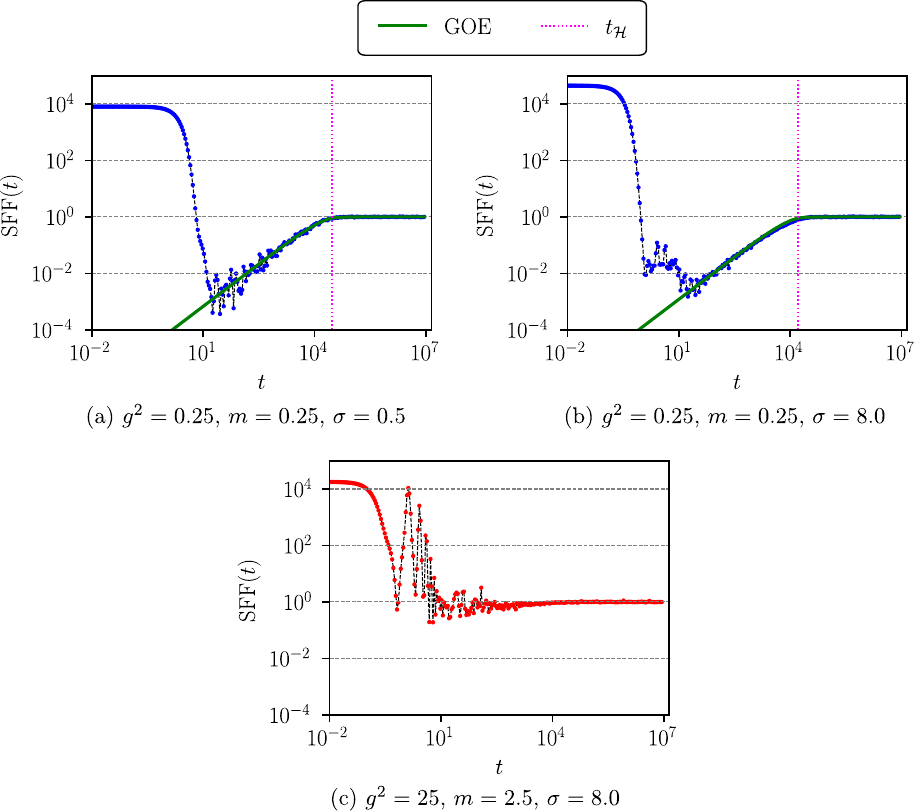}
\caption{The SFF for $N=14$ and $\jmax=1/2$ averaged over small time windows for a Gaussian filter function $f(E)$ with mean around the spectrum peak energy and width $\sigma$ is shown here for various Hamiltonian parameters and $\sigma$s.
In (a) and (b) we consider parameters for which the Hamiltonian in Eq.~\eqref{eq: complete Hamiltonian in KS} is chaotic, i.e., $g^2=0.25$ and $m=0.25$, while in (c) we take $g^2=25$ and $m=2.5$ for which the Hamiltonian is nearly integrable.
The SFFs in (a) and (b) have clear slope-ramp-plateau structures, where the ramp-plateau structure is consistent with the GOE prediction for the connected SFF given in Eq.~\eqref{eq: SFF_prediction}, shown as a green line. The ramp-plateau transition is denoted by $t_{\mathcal{H}}$ (dotted pink line). This structure is absent for the nearly integrable Hamiltonian in (c).
The fluctuations in the region before the slope starts in (a) and (b) are determined by different $\sigma$s, $\sigma=0.5$ and $\sigma=8$, respectively, where narrower filter in the former case suppresses the fluctuations in the slope region.
\label{fig: SFF_plots}
}
\end{figure}
Besides the distribution of level spacings, the spectral form factor (SFF) is another widely used tool to study chaotic behavior. It also encodes information of the approach to thermalization in real-time evolution of many different quantum systems~\cite{Haake:2010fgh,Mehta}. 
The SFF is defined as the ensemble-averaged squared magnitude of the Fourier transformed two-point correlation function of the density of states $\rho(E) = \sum_{a} \delta(E - E_a)$ with a suitable filter function $f(E)$~\cite{Winer:2020gdp}:
\begin{subequations}
    \begin{align}
        Z(t,f)&\equiv \sum_a f(E_a)e^{-iE_at} \\
        \text{SFF}(t, f) &=
        \overline{Z(t,f)Z^*(t,f)}
        =  \overline{\left|\int_{-\infty}^{\infty} dE f(E) \rho(E) e^{-iEt} \right|^2} , \label{eq:SFF}
    \end{align}
\end{subequations}
where $t$ denotes time, and the relevance of the filter function $f(E)$ will be discussed below. The overline indicates the ensemble average.
For a RMT ensemble, the properties of the SFF are known analytically~\cite{Winer:2020gdp, Barney:2023idq, Cotler:2016fpe, Cotler:2017jue, Liu:2018hlr, Saad:2018bqo}. Two important time scales that arise in the RMT study are the Thouless time $t_{\mathcal{T}}$ and the Heisenberg time $t_{\mathcal{H}}$. The former is defined as the time when the universal RMT behavior starts. The latter is defined with the expression
\begin{equation}
    t_{\mathcal{H}}= \frac{2 \pi \mathcal{N}_{\rm plateau}}{\int dE f(E)^2}
    \label{eq: Heisenberg time},
\end{equation}
where $\mathcal{N}_{\rm plateau}=\sum_a f(E_a)^2$.
When the filter function $f(E)$ is one, this expression reduces to the more familiar result set by the mean energy spacing, $t_{\mathcal{H}}=2\pi\mathcal{D}/(E_{\mathcal{D}-1} - E_0)$.
As a function of $t$, three different parts of the SFF are identified as follows: $(i)$ the ``slope" with added oscillations for the period before the Thouless time, i.e., $t\leq t_{\mathcal{T}}$, $(ii)$ the ``ramp" for $t$ between $t_{\mathcal{T}}$ and the Heisenberg time, i.e., $t_{\mathcal{T}}\leq t \leq t_{\mathcal{H}}$, and $(iii)$ the ``plateau" for the period after $t_{\mathcal{H}}$, i.e., $t_{\mathcal{H}}\leq t$.
The origin of the oscillations in the slope region can be traced to Wigner's semi-circle law for the spectrum~\cite{Das:2025pwy}, which asserts that the density of states for a RMT ensemble vanishes abruptly with discontinuities in its derivative.
These discontinuities generate the oscillations under the Fourier transform in Eq.~(\ref{eq:SFF}).
By introducing a suitable smooth but otherwise arbitrary smearing function $f(E)$ that acts as a low pass filter, one can try to suppress the RMT oscillations of the disconnected SFF.
Once the random nature starts to dominate, the ramp part of the SFF for a GOE begins with an approximate linear dependence on $t$. 
For a finite system, the ramp ends when it reaches its Heisenberg time. Beyond this point, the off-diagonal terms in Eq.~\eqref{eq:SFF} average to zero, and the SFF flattens to a plateau of the height $\mathcal{N}_{\rm plateau}$.
In the following, we will look at the normalized quantity $\mathrm{SFF}(t,f) \rightarrow \mathrm{SFF}(t,f)/\mathcal{N}_{\rm plateau}$, such that the plateau height is set to one. Furthermore, we consider a Gaussian filter function $f(E)=e^{(E-\mu_E)^2/(2 \sigma^2)}$ that is centered around the energy $\mu_E$ with width $\sigma$.

In Fig.~\ref{fig: SFF_plots}, results for the SFFs of the LSH Hamiltonian with $N=14$ and $\jmax=1/2$ are shown on a log-log scale for three different scenarios.
In order to mimic an ensemble average, we calculate for $10^6$ time steps that are uniformly distributed on the log scale and average over 3333 nearby data points. 
We then compare the ramp and the plateau parts of the results to the GOE predictions that originate from the ``connected" SFF defined as
\begin{align}
    {\rm SFF}^{\rm c}(t,f) &= \overline{ (Z(t,f)- \overline{Z(t,f) })(Z(t,f)- \overline{Z(t,f) })^*}.
\end{align}
The GOE prediction for the connected part of the SFF is given by~\cite{Winer:2020gdp, Cipolloni_2023}:
\begin{align}
    \mathrm{SFF}^{\mathrm{c}}(t)=\left\{\begin{array}{ll} \frac{2t}{t_{\mathcal{H}}}-\frac{t}{t_{\mathcal{H}}}\mathrm{log} \left(1+\frac{2t}{t_{\mathcal{H}}}  \right), &  0<t\leq t_{\mathcal{H}}, \\
    2-\frac{t}{t_{\mathcal{H}}} \mathrm{log}\left( \frac{2t+t_{\mathcal{H}}}{2t-t_{\mathcal{H}}} \right), & t\geq t_{\mathcal{H}},\end{array}\right.
    \label{eq: SFF_prediction}%
\end{align}
where $t_{\mathcal{H}}$ is evaluated using~\eqref{eq: Heisenberg time}.

We see in cases (a) and (b) that the SFF nicely follows the GOE prediction for the connected SFF in Eq.~\eqref{eq: SFF_prediction} after a certain $t$, which is the Thouless time $t_{\mathcal{T}}$,  approximately given by $t_{\mathcal{T}}=21$.
Before the universal ramp behavior starts, we can recognize non-universal oscillations that are strongly suppressed for a smaller $\sigma$, i.e., in (a), where only the eigenvalues around $\mu_E$ play a role and eigenvalues near the edges do not contribute significantly. 
In (c), the SFF calculated from the LSH Hamiltonian does not have a ramp region, nor exhibit the GOE behavior.
We attribute this to the breakdown of quantum ergodicity as indicated by the value of $\langle r \rangle=0.4188$, which indicates that the Hamiltonian is closer to an integrable system. Thus, we see that the SFF can also be used as an indicator of quantum chaos, as pointed out in Ref.~\cite{Winer:2020gdp}. Furthermore, the saturation plateau is approached from the above in (c). 
Details of the approach are sensitive to the approximate symmetries close to the integrable point, as recently explored in~\cite{Baumgartner:2024orz}.

\subsection{Operator ME fluctuations
\label{subsec: operator ME fluctuations}
}
The results in the previous subsection identify the parameter region that exhibits quantum chaos.
In the remaining section, we will verify and study the predictions of the ETH ansatz given in Eq.~\eqref{eq: ETH operator ansatz} for certain operators specified below.
For the following results in this section, we use the Hamiltonian parameters $g^2=0.25$ and $m=0.25$ from the chaotic region of the parameters space, and only vary $N$ and $\jmax$, unless mentioned otherwise.
This subsection checks the smoothness of $\langle E_a|O|E_a\rangle$ as a function of $E_a/N$, its value compared to the microcanonical value of $O_{\rm mc}(E_a)$, and the dependence of variances in the diagonal and off-diagonal parts on system sizes.

The operators considered in this work are gauge-invariant Hermitian operators which are either local (including nearest neighbor interactions) or extended.
The local operators, $h_I(x,x+1)$, $h_E(x)$, and $h_M(x)$ are parts of the Hamiltonian as defined in Eqs.~\eqref{eq: hM in LSH}-\eqref{eq: hI in LSH}.
The extended operator that we consider here is referred to as the Hermitianized string operator of length $L$. Mathematically it is given by
\begin{equation}
    S'_L(x)  =   \psi^\dagger(x)U(x,x+1)U(x+1,x+2)\cdots U(x+L-1,x+L)\psi(x+L) + {\rm H. c.},
    \label{eq: string operator def}
\end{equation}
where the contracted color indices are suppressed for brevity, and H.c.~indicates the Hermitian conjugate of the first term on the right hand side.
The LSH expression for $S'_L(x)$ can be obtained by substituting the Schwinger boson decomposition of $U(x,x+1)$ from Eqs.~\eqref{eq: U in UL and UR}-\eqref{eq: UL and UR in schwinger bosons} into Eq.~\eqref{eq: string operator def} and identifying the gauge invariant operators from Table~\ref{tab: LSH operators}:
\begin{align}
    S'_L(x)  =   &\frac{1}{\sqrt{N_L(x)+1}} 
    \begin{pmatrix}
        \mathcal{S}^{++}_{\rm out} \\  \mathcal{S}^{+-}_{\rm out}
    \end{pmatrix}^T_x
    \frac{1}{\sqrt{N_L(x+1)+1}}
    \begin{pmatrix}
        \mathcal{L}^{++} & \mathcal{L}^{+-}\\
        -\mathcal{L}^{-+} & \mathcal{L}^{--}
    \end{pmatrix}_{x+1}
    \frac{1}{\sqrt{N_R(x)+1}}\,\times\nonumber\\
    \cdots \times\,     
    &\frac{1}{\sqrt{N_L(x+L-1)+1}}
    \begin{pmatrix}
        \mathcal{L}^{++} & \mathcal{L}^{+-}\\
        -\mathcal{L}^{-+} & \mathcal{L}^{--}
    \end{pmatrix}_{x+L-1}
    \frac{1}{\sqrt{N_R(x+L-1)+1}}\nonumber\\
    &\times\begin{pmatrix}
        \mathcal{S}^{+-}_{\rm in}\\ \mathcal{S}^{--}_{\rm in}
    \end{pmatrix}_{x+L}
    \frac{1}{\sqrt{N_R(x)+1}}
    + {\rm H. c.},
    \label{eq: string operator in LSH}%
\end{align}
where the superscript $T$ means transpose, the subscripts on the matrices denote the common lattice site label of its operator entries, and the expressions for $\mathcal{S}$ and $\mathcal{L}$ operators in terms of LSH operators are given in Table~\ref{tab: LSH operators}.
The string operator is considered extended for $L>1$ since comparing Eq.~\eqref{eq: string operator def} to Eq.~\eqref{eq: HI in KS} implies that $S'_1(x) = h_I(x,x+1)$, which is a local operator.
\begin{figure*}[t!]
    \centering
    \includegraphics[scale=1.0]{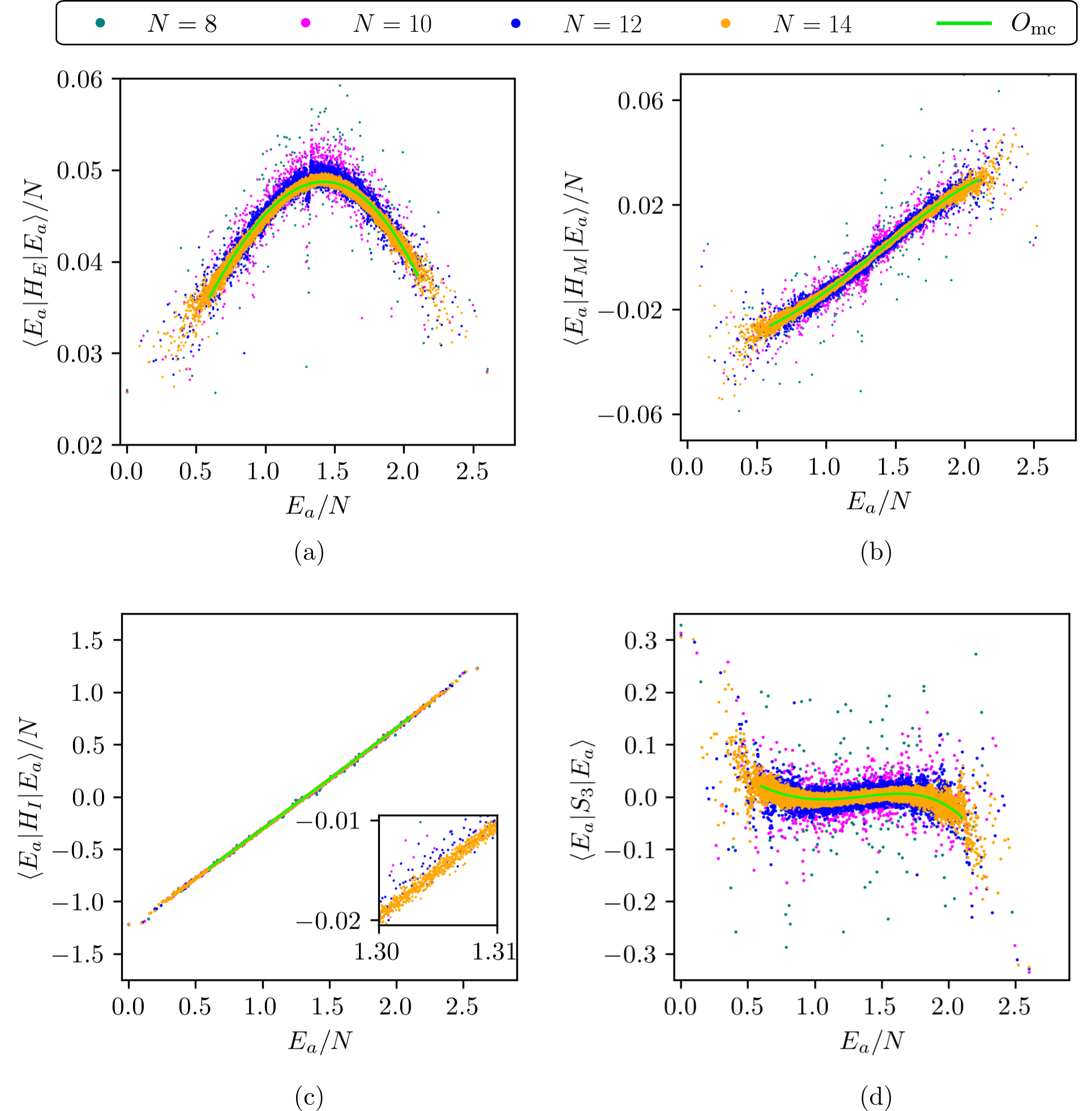}
    \caption{The diagonal MEs $\langle E_a | O |E_a \rangle$ of the observables $O=H_E/N$, $H_M/N$, $H_I/N$, and $S_3$ for the system sizes $N=8$, $10$, $12$ and $14$ are plotted against the energy density\footnote{We rescale the x-axis with $N$ so that the spectrum ranges of different lattices roughly fit in the same range.}. The microcanonical value fitted by a degree-four polynomial, is indicated by the green line. The fit values of the coefficients obtained for $N=14$ are listed in Table~\ref{tab: diagonal ME polynomial fit}. The fluctuations around the microcanonical value decrease in the thermodynamic limit for all operators shown here. The size of fluctuations compared to their central value is very small for $O=H_I/N$, and thus, not visible in (c). The inset plot in (c) shows a magnified region to illustrate this.
    \label{fig: diagonal ME}
    }
\end{figure*}

The MEs of an operator between two zero momentum states is given by the translationally invariant component of the operator, since the momentum transfer is forced to be zero.
We construct the translationally invariant operators by averaging over all lattice sites.
For the local operators $h_M(x,x+1)$, $h_E(x)$ and $h_I(x)$, this results in $H_M/N$, $H_E/N$, and $H_I/N$, respectively, and for the extended string operator, it leads to
\begin{equation}
    S_L(x) = \frac{1}{N} \sum_{x=0}^{N-1} S'_L(x).
    \label{eq: string operator zero momentum def}
\end{equation}
With these, we test the predictions of the ETH ansatz in Eq.~\eqref{eq: ETH operator ansatz} for local operators $O\in \{ H_M/N, H_E/N, H_I/N\}$ and extended operators $O=S_L$ for a few selected values of $L$.
\begin{table}[t!]
    \renewcommand{\arraystretch}{1.5}
    \begin{center}
    \begin{tabular}{C{2cm}| C{2cm} C{2cm} C{2cm} C{2cm} C{2cm}}
     $O$ & $a_0$ & $a_1$ & $a_2$ & $a_3$ & $a_4$ \\
    \hline
    \hline
    $H_E/N$ & 0.0169 & 0.0341 & 0.0011 & -0.0080 &  0.0010\\
    $H_M/N$ & -0.0362 &  0.0077 &  0.0131 & 0.0050 & -0.0028\\
    $H_I/N$ & -1.2029 & 0.7793 & 0.2075 & -0.1126 & 0.0234 \\
    $S_3$ & 0.1359 & -0.2495 & 0.0502 & 0.0998 & -0.0410 \\
    \hline
    \end{tabular}
    \end{center}\caption{Fitted coefficients of the fourth degree polynomial $\sum^4_{i=0} a_ix^i$ for the diagonal MEs of operator $O$. Here $a_i$ denotes the coefficient of the monomial of order $i$ and $x_i$ represents the variable $E_a/N$. The range of fitting is taken over the whole energy region, and the fitting errors are small and thus are not shown here.
    \label{tab: diagonal ME polynomial fit}
    }
\end{table}

In Fig.~\ref{fig: diagonal ME}, we plot the diagonal MEs of $O$ in the energy eigenbasis against the corresponding energy eigenvalues rescaled with $N$ for four different lattice sizes: $N=$ 8, 10, 12, and 14.
Here, we considered the LSH Hamiltonian with local Hilbert space truncation $j_{\rm max} = 1/2$.
As can be seen, the variances of the diagonal MEs shrink as the lattice size increases.
This is most manifest in the case of $H_E/N$, $H_M/N$ and $S_3$, where the variances are about one order of magnitude smaller than the central values.
In the case of $H_I/N$, the variances are much smaller than the ME values, and thus, they are not visible in the plot.
This is due to the choice of the Hamiltonian parameters, $g^2=0.25$ and $m=0.25$, which results in energy eigenvalues receiving suppressed contributions from $H_E$ and $H_M$ relative to the contribution from $H_I$.
As the sum of deviations from the mean values in the diagonal MEs of these three operators vanish in the energy eigenbasis, the deviations in diagonal MEs of $H_I$ are equal in magnitude to the sum of deviations in diagonal MEs of $H_E$ and $H_M$.
However, the central values of the diagonal MEs of $H_I$ are much larger compared to their deviations, and thus, the variances are only visible in the inset plot.
The narrowing of the diagonal ME bands as the system size increases indicates that for all operators the values of the diagonal MEs fluctuate around a smooth curve representing the microcanonical values of the corresponding operator, $O_{\rm mc}$.
We obtain this curve by fitting a polynomial of degree four to the narrowest band given by $N=14$.
The fitted coefficients are tabulated in Table~\ref{tab: diagonal ME polynomial fit}, and the respective polynomial functions are denoted with solid green lines in Fig.~\ref{fig: diagonal ME}.
\begin{figure*}[t!]
    \centering
    \includegraphics[scale=1.0]{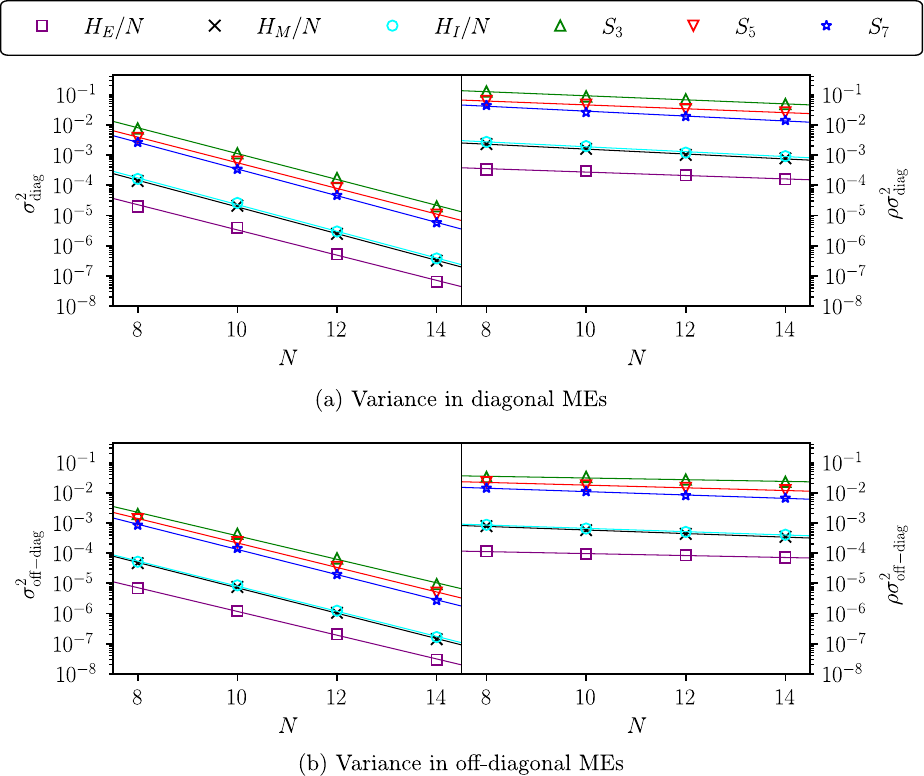}
    \caption{The variances in diagonal, $\sigma^2_{\mathrm{diag}}$, and off-diagonal, $\sigma^2_{\mathrm{off-diag}}$, operator MEs are plotted against $N$ on the left column in (a) and  (b), respectively, for three local ($O=H_E/N$, $H_M/N$, $H_I/N$) and three extended ($O=S_3$, $S_5$, $S_7$) operators. The truncation is $\jmax=1/2$ for all cases. Both variances decrease exponentially with $N$, as indicated by the solid lines, which are obtained by fitting with an exponential function $a\,e^{-bN}$. In a similar fashion, the column on the right shows the same variances on the left but multiplied with the density of states $\rho$ for each $N$. Thus, the plots on the right denote that $f_{O}^2(\omega \approx 0)$ has insignificant dependence on $N$. 
    The values $a$ and $b$ obtained from all fits are shown in Table~\ref{tab: exponential fit variances}.
    This dependence on $N$ in both diagonal and off-diagonal variances is consistent with ETH predictions.
    \label{fig: off diagonal ME variance}
    }
\end{figure*}

According to the ETH, the variances in the diagonal as well as the off-diagonal parts of the MEs are expected to be exponentially suppressed in the system size by the factor $e^{-S(\bar{E})/2}$.
This is verified in the left column in Fig.~\ref{fig: off diagonal ME variance}, where we have plotted the variances in the diagonal and off-diagonal MEs, i.e., $\sigma^2_{\rm diag}$ and $\sigma^2_{\rm off-diag}$, for four different lattice sizes.
The former variance is calculated with the mean value obtained from the polynomial fit, while the latter variance is calculated by taking a zero mean as the off-diagonal MEs average to zero in a small energy window.
Furthermore, we restrict ourselves to an energy range of $E_a$ that satisfies Eq.~\eqref{eq: energy window def} to avoid the edge effects of the spectrum and consider MEs with $\omega<0.1$, which is two orders of magnitude smaller than the mean eigenenergy.
Both variances are fitted by a function $a\,e^{-bN}$ and the fitted parameter values are listed in Table~\ref{tab: exponential fit variances}. The fitting results are depicted as solid lines in the left panels in Fig.~\ref{fig: off diagonal ME variance}, showing that both $\sigma^2_{\rm diag}$ and $\sigma^2_{\rm off-diag}$ for an array of operators considered here fall exponentially fast in $N$, which is consistent with the ETH prediction.
This indicates that the second term in Eq.~\eqref{eq: ETH operator ansatz} falls off exponentially fast in the thermodynamic limit with MEs converging to the microcanonical values.
Moreover, the rate of exponential decay appears to be independent of the operator, an observation that is consistent with the ETH prediction which states that the rate is predominantly determined by the entropy density averaged over the energy range considered, see Eq.~\eqref{eq: ETH operator ansatz}.
This is further verified by the very close values of the parameter $b$ as listed in Table~\ref{tab: exponential fit variances}.
Finally, we compare $\sigma^2_{\rm diag}$ and $\sigma^2_{\rm off-diag}$ with $\rho\sigma^2_{\rm diag}$ and $\rho\sigma^2_{\rm off-diag}$ that are plotted in the right column in Fig.~\ref{fig: off diagonal ME variance}. 
Here $\rho$ denotes the density of states for each $N$, and thus, the right column denotes the dependence of $f_{O}^2(\omega \approx 0)$ on $N$.
It is clear from Fig.~\ref{fig: off diagonal ME variance} that the function $f_O^2(\omega \approx 0)$ has milder dependence on $N$ for all operators.
This is further demonstrated by fitting the plots on the right in a similar manner to the plots on the left. The fitted parameter values are listed in Table~\ref{tab: exponential fit variances}.
\begin{table}[t]
    \renewcommand{\arraystretch}{1.3}
    \begin{center}
    \begin{tabular}{C{1.5cm}| C{1.5cm} C{1.5cm} | C{1.5cm} C{1.5cm} |C{1.5cm} C{1.5cm} | C{1.5cm} C{1.5cm}}
     \multirow{2}{*}{$O$} & \multicolumn{2}{c|}{$\sigma^2_{\rm diag}$} & \multicolumn{2}{c|}{$\sigma^2_{\rm off-diag}$} & \multicolumn{2}{c|}{$\rho\sigma^2_{\rm diag}$} & \multicolumn{2}{c}{$\rho\sigma^2_{\rm off-diag}$} \\
     \cline{2-9}
     & $a$ & $b$ & $a$ & $b$  & $a$ & $b$ & $a$ & $b$ \\ 
    \hline
    \hline
    
    $H_E/N$ & 0.0500  & 0.9619 & 0.0101 & 0.9059 & 0.0010 & 0.1308 & 0.002 & 0.0748\\
    $H_M/N$ & 0.5062 & 1.0182 & 0.1124 & 0.9673 & 0.0103 & 0.1872 & 0.0023 & 0.1363  \\
    $H_I/N$ & 0.6052 & 1.0182 & 0.1162 & 0.9580 & 0.0123 & 0.1872 & 0.0024 & 0.1269\\
    $S_3$ & 21.8357 & 0.9859 & 2.9141 & 0.8963 & 0.4329 & 0.1548 & 0.0590 & 0.0652 \\
    $S_5$ & 9.9177 & 0.9787 & 2.4676 & 0.9340 & 0.2008 & 0.1476 & 0.0500 & 0.1030 \\
    $S_7$ & 9.1679 & 1.0184 & 1.9585 & 0.9602 & 0.1856 & 0.1874 & 0.0396 & 0.1292 \\
    \hline
    \end{tabular}
    \end{center}
    \caption{Values of the parameters $a$ and $b$ in $a\, e^{-bN}$ when fitted to $\sigma^2_{\rm diag}$, $\sigma^2_{\rm off-diag}$, $\rho\sigma^2_{\rm diag}$, and $\rho\sigma^2_{\rm off-diag}$ in Fig.~\ref{fig: off diagonal ME variance} are shown here for each operator. The physical meaning of the parameter $b$ in the first two columns is primarily the averaged entropy density. The values of $b$ in the last two columns indicate that $f^2(\bar{E},\omega)$ has insignificant dependence on $N$. The errors in the fit values of $b$ are less than 5\%, and thus are not shown here.
    \label{tab: exponential fit variances}
    }
\end{table}
\begin{figure*}[t!]
    \centering
    \includegraphics[scale=1.0]{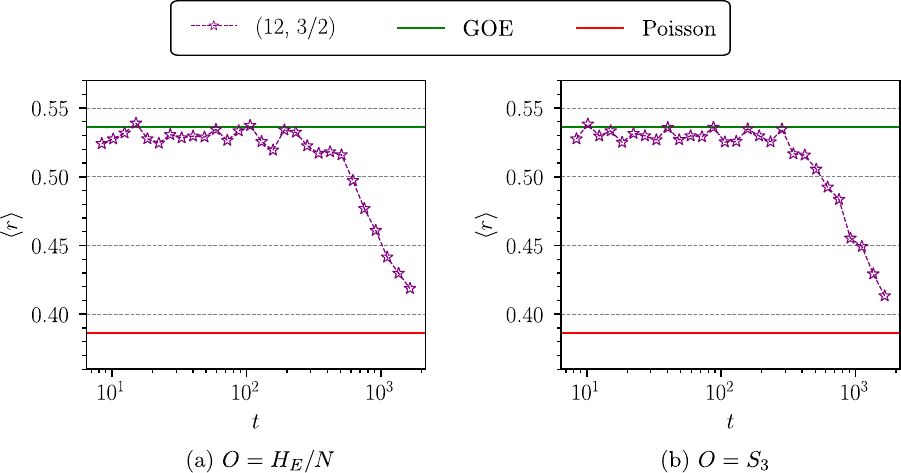}
    \caption{The $\langle r\rangle$ values of band matrices $\tilde{O}^{(t)}$ for both $H_E/N$ and $S_3$ operators with $(N,\jmax)=(12,3/2)$.
    An agreement with the GOE value is approximately attained even for small time $t\lesssim10$, while deviations become significant at large time $t\gtrsim 500$.
    \label{fig: restricted gap ratio for band matrix}
    }
\end{figure*}

\subsection{Randomness
\label{subsec: randomness}
}

In this subsection, we test when the correction to the microcanonical ensemble expectation value, i.e., the $R_{ab}f_O(\bar{E},\omega)$ term in the ETH ansatz~\eqref{eq: ETH operator ansatz} shows GOE behavior.
To this end, we look at different RMT measures, where predictions for a GOE exist, and compare them to the results obtained from our physical system. 
We use one local and one extended observable, namely $H_E/N$ and $S_3$, respectively, to carry out this analysis.

In the following, we define a band matrix of an operator $O$ in the energy window of size $\Delta E$ centered around the energy $E$ as:
\begin{align}
    \label{eq:O^T}
    \tilde{O}_{ab}^{(t)} &= \left\{
    \begin{array}{ll}
    \langle E_a |O| E_a \rangle - O_{\rm mc}(E_a) &\ E_a=E_b, \\
    \langle E_a |O| E_b \rangle &\ 0<|E_a-E_b| \leq \frac{2\pi}{t}, \\
    0 &\
    \frac{2\pi}{t} < |E_a - E_b | , 
    \end{array} \right. \,
\end{align}
where $E_a$, $E_b\in [E-\Delta E/2, E+\Delta E/2 ]$ and the bandwidth of $\tilde{O}^{(t)}$ is given by $2\pi/t$ where $t$ can be interpreted as time, controlling the range of contributing frequencies.
All of the following results in this subsection are calculated by taking the band matrix centered at the energy $E=17.67*N/12$ with $\Delta E=0.75$, unless explicitly stated otherwise. We note that with increasing time $t\gg \pi/\Delta E$, there are fewer non-zero MEs, hence statistical errors increase.
\begin{figure*}[t!]
    \includegraphics[scale=1.0]{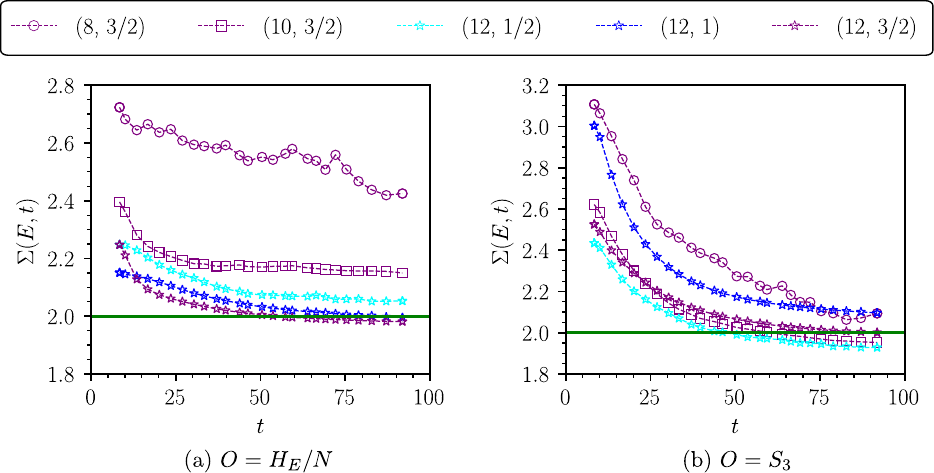}
    \caption{The $\Sigma(E,t)$ measure defined in Eq.~\eqref{eq: Sigma definition} is shown for two operators $O=H_E/N$ and $O=S_3$ in (a) and (b), respectively. Different combinations of $N$ and $\jmax$ are considered and are denoted in the legends as $(N,\jmax)$. The GOE prediction for the value of $\Sigma(E,t)$ is two, which is indicated by the green line. Each shape (color) identifies a value of $N$ $(\jmax)$.
    We see that $\Sigma(E,t)$ approaches the GOE prediction for increasing values of $N$ and $\jmax$ only at $t\gtrsim 50$, unlike in Fig.~\ref{fig: restricted gap ratio for band matrix}.
    \label{fig: Sigma_size_and_jmax_dependence}
    }
\end{figure*}

As stated in Sec.~\ref{subsec: ETH}, the value of $\langle r \rangle$ obtained from the eigenvalues of a matrix is a good indicator to determine whether the matrix obeys GOE or Poisson statistics.
For the band matrix $\tilde{O}^{(t)}$, the values of $\langle r \rangle$ are plotted in Fig.~\ref{fig: restricted gap ratio for band matrix} for varying band sizes given by different values of $t$.
We observe that $\tilde{O}^{(t)}$ obeys GOE statistics even at small time ($t\lesssim10$). Deviations start to grow significant at late time ($t\gtrsim 500$), which we attribute to the fact that the eigenvalues of band matrices with a sufficiently small width become uncorrelated~\cite{Richter:2020bkf}. 
Therefore, the RMT measures are not reliable for late time. This is an effect of the method rather than the physical theory.
However, $\langle r \rangle$ is not a sufficient marker for the complete GOE behavior as there could exist ME correlations that only show up in other GOE measures to be discussed below.

The first such marker is related to the variances in the diagonal and off-diagonal MEs.
The GOE predicts that the former variance is twice the latter variance.
Thus, the ratio of these two variances for the band matrix $\tilde{O}^{(t)}$, defined as
\begin{align}
    \Sigma({E},t) \equiv \frac{\sigma_{\rm diag}^2({E})}{\sigma_{\rm off-diag}^2({E},t)} \,,
    \label{eq: Sigma definition}
\end{align}
where $\sigma_{\rm off-diag}^2({E},t)$ is calculated over off-diagonal entries inside the band, is expected to approach 2 as the theory exhibits GOE behavior.
The results of $\Sigma(E,t)$ for $O = H_E/N$ and $S_3$ are shown in Fig.~\ref{fig: Sigma_size_and_jmax_dependence}.
We studied the effects of increasing lattice sizes and local Hilbert space truncations as shown in the figure. 
We find that for a fixed lattice size $N=12$, the $\Sigma(E,t)$ measure approaches the GOE prediction with increasing values of $\jmax$.
Furthermore, for a fixed $\jmax=3/2$, we see a similar convergence in the thermodynamic limit.
In both cases, the agreement occurs at late time $t\gtrsim50$.

\begin{figure*}[t!]
    \includegraphics[scale=1.0]{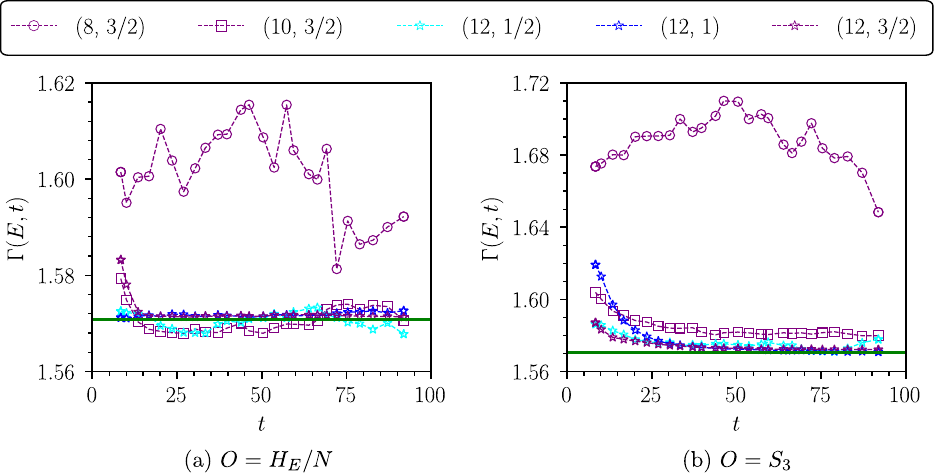}
    \caption{The $\Gamma(t)$ measure defined in Eq.~\eqref{eq: Gamma measure definition} is shown for different combinations of $(N,\jmax)$ for the observable $H_E/N$ and $S_3$. The solid green line indicates the GOE prediction $\pi/2$, which is approached with increasing $N$ and $\jmax$ at late times, as in Fig.~\ref{fig: Sigma_size_and_jmax_dependence}.
    \label{fig: Gamma_size_and_jmax_dependence}
    }
\end{figure*}

Another indicator is the $\Gamma$ measure defined~\cite{LeBlond:2019eoe} as 
\begin{align}
    \Gamma({E}, t) \equiv \frac{\sigma_{\rm off-diag}^2({E},t)}{|\tilde{O}_{\rm off-diag}|_{\rm avg}^2(E,t)},
    \label{eq: Gamma measure definition}
\end{align}
where $|\tilde{O}_{\rm off-diag}|_{\rm avg}$ is the average of absolute values of off-diagonal entries of $\tilde{O}^{(t)}$ inside the band and depends on $E$ and $t$.
For GOE $\Gamma=\pi/2$.
We have plotted the $\Gamma$ measure for the same parameters and operators as in Fig.~\ref{fig: Sigma_size_and_jmax_dependence} in Fig.~\ref{fig: Gamma_size_and_jmax_dependence}.
Again, for the largest system, the agreement of $\Gamma$ with its GOE prediction is attained at late times. For both the $\Gamma$ and $\Sigma$ measures, the deviation of $(8,3/2)$ from the GOE prediction is relatively large compared to $N=10$ and $N=12$. This could be caused by the limited statistics for this system size, namely $81$ states, in this energy window $\Delta E = 0.75$. For $(10,3/2)$ and $(12,3/2)$, we have $589$ and $4049$ states in the same energy window.

The $\Sigma$ and $\Gamma$ measures are only sensitive to the magnitudes of the MEs relative to the microcanonical ensemble expectation values.
Now we will study two measures that are sensitive to not only the magnitudes but also the signs.
The first measure, $\Lambda_n$, for the band matrix $\tilde{O}^{(t)}$ is defined by considering the $n$-th moment of the matrix $M_n(\tilde{O}^{(t)})$
\begin{align}
    \Lambda_n(E,t) =\frac{[M_n({\tilde{O}^{(t)}})]^{\frac{n+2}{n}}}{M_{n+2}({\tilde{O}^{(t)}})},
    \label{eq: Lambda n definition}
\end{align}
where the dependence on $E$ and $t$ is through the band matrix.
The matrix moments are defined as
\begin{align}
    M_n({\tilde{O}^{(t)}})=\frac{1}{d}\mathrm{Tr}[(\tilde{O}^{(t)})^n],
    \label{eq: Mn moment definition}
\end{align}
where $d$ is the dimension of the band matrix. 
For the case of $n=2$, the $\Lambda_2(E,t)$ measure has been studied previously to determine the onset of GOE behavior~\cite{Wang:2021mtp,Ebner:2023ixq}.
In this work, we consider both $n=2$ and higher (even) moments up to $n=10$. 
The $n$-th moment for the GOE is given by $M_{2n}=C_n \sigma^{2n}$ and $M_{2n+1}=0$, where $\sigma$ is the variance of the GOE and $C_n$ are the Catalan numbers defined as $C_n=\frac{(2n)!}{(n+1)!n!}$~\cite{anderson2010introduction}.
For manifest comparison, we will divide $\Lambda_n(E,t)$ calculated for our lattice system by its GOE prediction, which is
\begin{align}
    {\rm GOE}(\Lambda_n)=\frac{C_n^{\frac{n+2}{n}}}{C_{n+1}}.
    \label{eq: Lambda n measure GOE prediction}
\end{align}
\begin{figure*}[t!]
    \centering
    \includegraphics[scale=1.0]{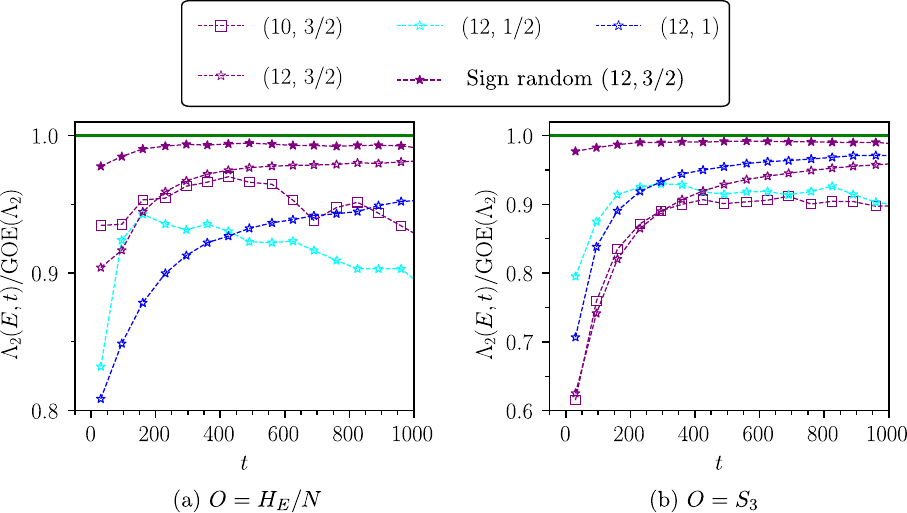}
    \caption{The $\Lambda_2(E,t)$ measure defined in Eq.~\eqref{eq: Lambda n definition} for operators $H_E/N$ in (a) and $S_3$ in (b) for different combinations of $(N,\jmax)$.
    The $y$-axis is scaled by the GOE prediction for $\Lambda_2$ given in Eq.~\eqref{eq: Lambda n measure GOE prediction}. Thus, the green line at 1.0 indicates complete agreement with GOE behavior. 
    We see large deviation from the GOE behavior at late times if the statistics is limited, either because $N$ or $\jmax$ are not large enough. 
    For sufficiently large $N$ and $\jmax$, say, $(12,3/2)$, the $\Lambda_2(E,t)$ measure approaches the GOE prediction as time increases, though deviation still exists.
    Furthermore, the measure for $(12, 3/2)$ is compared to its sign-random version, indicating that the deviation from GOE behavior for $200\lesssim t \lesssim 1000$ is primarily a result of sign correlations in the MEs. For $t>1000$, the eigenvalues of the band matrices exhibit Poisson statistics, as shown in Fig.~\ref{fig: restricted gap ratio for band matrix}.
    \label{fig: Lambda_2_measure}
    }
\end{figure*}

\begin{figure*}[h!]
    \centering
    \includegraphics[scale=1.0]{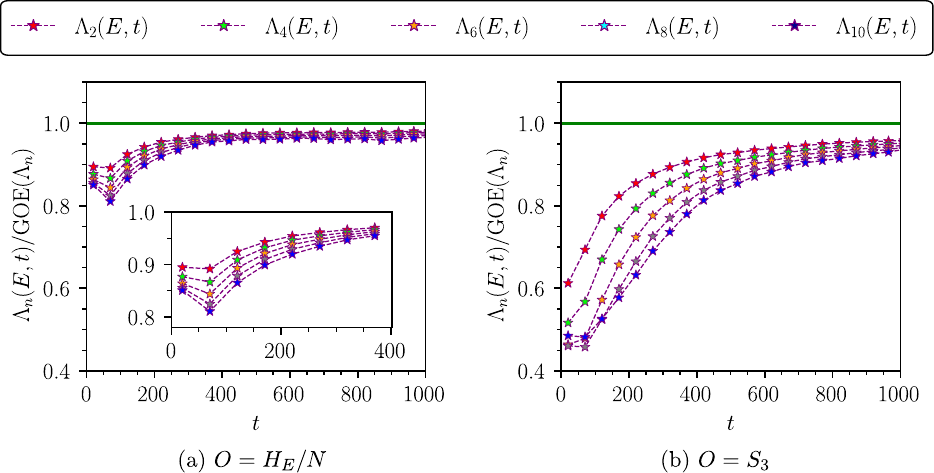}
    \caption{The $\Lambda_n(E,t)$ measure scaled by its GOE prediction is shown for the observables $H_E/N$ and $S_3$ for different ratios of moments specified by $n$, see Eq.~\eqref{eq: Gamma measure definition}. 
    The values of $(N,\jmax)$ are fixed at $(12, 3/2)$.
    The green line at 1.0 denotes complete agreement with GOE.
    The inset in (a) shows a magnified region for a better visibility of the trend for different $\Lambda_n$s. Larger deviation is seen with increasing $n$, in particular for the extended operator $S_3$.\vspace{0.1cm}
    \label{fig: Lambda_n_measure}
    }
\end{figure*}

We show the results for $\Lambda_2(E,t)$ in Fig.~\ref{fig: Lambda_2_measure} for different $N$ and $\jmax$, and compare $\Lambda_2$ for $N=12$ and $\jmax=3/2$ with $\Lambda_n$ for the same parameters in Fig.~\ref{fig: Lambda_n_measure}.
In Fig.~\ref{fig: Lambda_2_measure}, we find that for system sizes $N=8$ and $10$, the $\Lambda_2$ measure significantly deviates from the GOE value even at late time $t$, which we attribute to the limited statistics.
A small system, even non-integrable, is not expected to fully develop the GOE behavior.
For $N=12$, as $t$ increases, the $\Lambda_2(E,t)$ values of both operators approach the GOE value but deviations remain.
At $200\lesssim t\lesssim 1000$, the deviation from the GOE value is mainly caused by the sign correlations in the band MEs.
This can be seen from the fact that the $\Sigma$ and $\Gamma$ measures, which are only sensitive to the magnitudes, obey the GOE behavior already at the time $t\sim 50$. 
A further demonstration of this can be seen from the $\Lambda_2$ measure of a sign-randomized version of the band matrix, where the signs of the MEs are randomly taken from $\{+,-\}$ while maintaining the matrix Hermiticity. 
This way, one can remove the sign correlations, and we see that in the time range $200\lesssim t\lesssim1000$ the $\Lambda_2(E,t)$ value of the sign-randomized version is much closer to the GOE value. 
At even larger time $t>1000$, the $\langle r \rangle$ value deviates from the GOE value significantly, as demonstrated in Fig.~\ref{fig: restricted gap ratio for band matrix}, and thus, it is not useful to study the $\Lambda_2(E,t)$ measure in that regime.
In Fig.~\ref{fig: Lambda_n_measure},
we observe that the $\Lambda_n$ measures for $n>2$ also follow the GOE predictions approximately at late time. 
On the other hand, at a fixed earlier time, say $t=100$, deviation from the GOE value is larger for bigger values of $n$, indicating the non-Gaussian nature of the band matrix.
Furthermore, the larger deviation from the GOE value for the extended operator $S_3$ imply that extended operators exhibit more non-Gaussian features in thermalization dynamics. In the future, one may use cumulants to study higher order correlations, since cumulants are independent of each other. Although higher order moments can capture higher order correlations, they depend on the lower order ones.
\begin{figure*}[t!]
    \centering
    \includegraphics[scale=1.0]{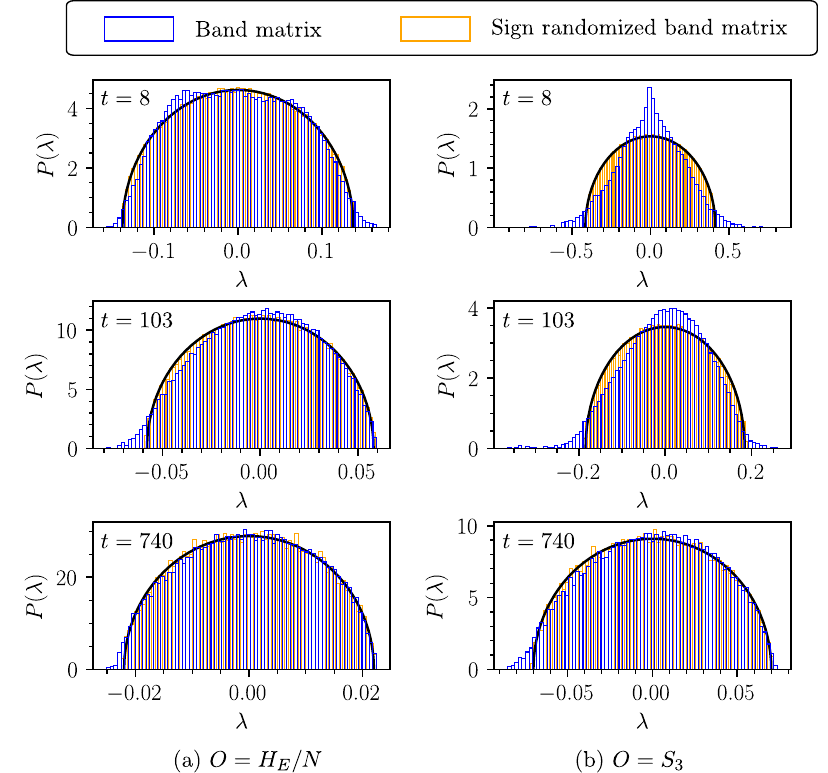}
    \caption{The probability distribution of eigenvalues of band matrices (blue) for observables $H_E/N$ and $S_3$ are shown in columns (a) and (b), respectively, where different rows denote different time: $t=8$ (top row), $t=103$ (central row) and $t=740$ (bottom row). Here, we fix $(N,\jmax)=(12,3/2)$. The GOE prediction for this distribution is given by the semicircle law (black line) as defined in Eq.~\eqref{eq: semicircle law}. We see agreement with the GOE behavior at late time, while the deviation at early time is primarily caused by the sign correlations in the band MEs. This is seen by the improved agreement with the semicircle law for the distribution of eigenvalues of sign-randomized band matrices (yellow). The early-time deviation is larger for the extended operator $S_3$, indicating a slower rate for complete thermalization.
    \label{fig:semicircle_N12_jmax1.5_HM_S5}
    }
\end{figure*}
\newpage
Finally, we test the distribution of eigenvalues of the band matrix $\tilde{O}^{(t)}$ for different values of $t$.
If the MEs are drawn from a GOE, the distribution of its eigenvalues is given by the semicircle law:
\begin{equation}
    P_t(\lambda)=\left\{\begin{array}{ll} \frac{1}{2 \pi M_2}\sqrt{4M_2-\lambda^2}, &\ |\lambda|\leq 2\sqrt{M_2}, \\
    0, & \ |\lambda|\geq 2\sqrt{M_2},\end{array}\right. ,
    \label{eq: semicircle law}
\end{equation}
where $M_2$ is the second moment of the band matrix $\tilde{O}^{(t)}$ as defined in Eq.~\eqref{eq: Mn moment definition} and $\lambda$ denotes the eigenvalues of the band matrix.
In Fig.~\ref{fig:semicircle_N12_jmax1.5_HM_S5}, we plot the eigenvalue distributions for both $H_E/N$ and $S_3$ operators at $N=12$ and $\jmax = 3/2$ and see that as $t$ increases, each distribution gets closer to a semi-circle given by Eq.~\eqref{eq: semicircle law}. 
We also see that the deviation from the GOE prediction is mainly caused by the sign correlations in the MEs, as the sign-randomized result is already close to the GOE semicircle at small values of $t$, where the distribution of the original band matrix has clear deviations. Furthermore, the deviation for the extended operator $S_3$ is larger than for $H_E/N$, indicating a slower thermalization rate for extended observables.
These observations are consistent with the conclusions drawn from the $\Lambda_n(E,t)$ analysis in Figs.~\ref{fig: Lambda_2_measure} and~\ref{fig: Lambda_n_measure}.

\subsection{$f_O$ function
\label{subsec: fO function}
}
\begin{figure}[t!] 
    \centering
    \includegraphics[scale=1.0]{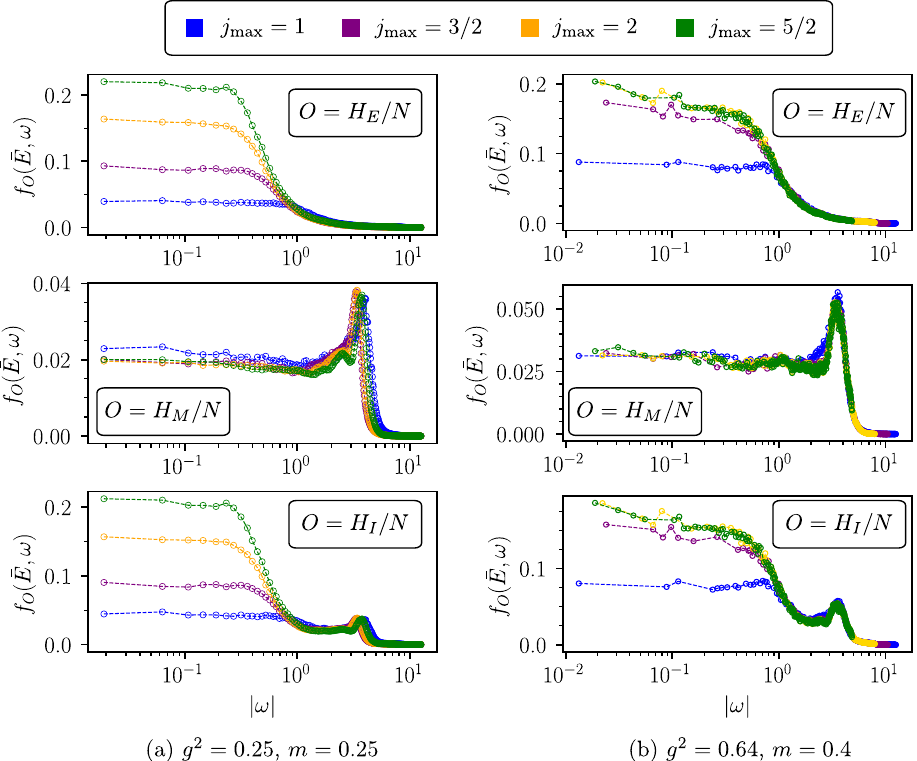}
    \caption{The function $f_O(\bar{E},\omega)$ is shown for local operators for two sets of parameter values: $g^2=0.25$, $m=0.25$ (a) and $g^2=0.64$, $m=0.4$ (b). The system size is fixed at $N=10$ and the central energy is taken as $\bar{E}=14.7417$ for (a) and $\bar{E}=14.1482$ for (b). The dependence of $f_O(\bar{E},\omega)$ on $\jmax$ is shown for all cases. The convergence in $\jmax$ is seen for all operators at $|\omega|\gtrsim 1$. In the $\omega\to0$ region, larger values $\jmax$ are required for the convergence of the $f_O$ functions for $H_E/N$ and $H_I/N$ in (a) while convergence is seen in all the other cases.
    \label{fig: LSH_f_function jmax dependence}}
\end{figure}
\begin{figure}[t] 
    \centering
    \includegraphics[scale=1.0]{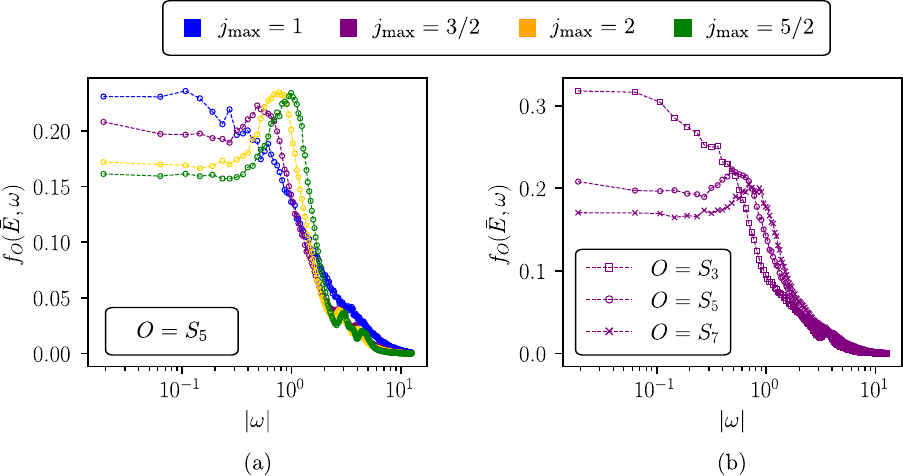}
    \caption{The function $f_O(\bar{E},\omega)$ is shown here for extended operators. The $\jmax$ dependence is shown in (a) for $S_5$ and the string length dependence is shown in (b) for $\jmax=3/2$. The system size is fixed at $N=10$ and the central energy is taken as $\bar{E}=14.7417$.
    \label{fig: LSH_f_function_string jmax dependence}
    }
\end{figure}
\begin{figure*}[t!]
    \centering
    \includegraphics[scale=1.0]{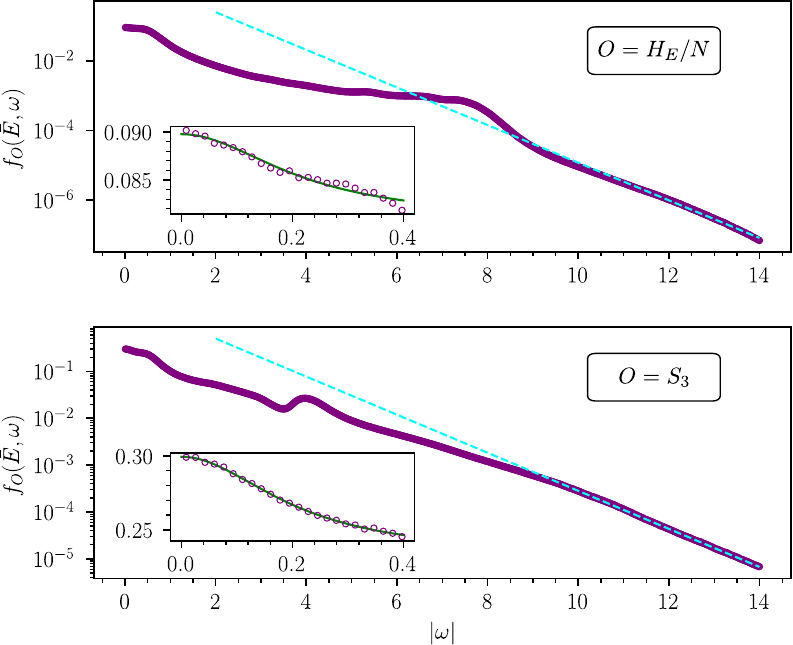}
    \caption{The $f_O$ function $f_O(\bar{E},\omega)$ is shown for observables $H_E/N$ and $S_3$ in the top and the bottom panel, respectively. We take $(N,\jmax)$ to be $(12,3/2)$ and choose the central energy $\bar{E}=17.67$. The small $|\omega|$ behavior is fitted with the Lorentzian shape given in Eq.~\eqref{eq: Lorentian for spectral function at small omega} and shown in the insets. At $|\omega|>11$, the $f_O$ function is fitted to an exponential function of the form $e^{-\beta' |\omega|/4}$. The fitted parameters are listed in Table~\ref{tab: Lorentzian fit and beta}. 
    \label{fig: f_function decay}
    }
\end{figure*}

The $f_O(\bar{E},\omega)$ function that appears in Eq.~\eqref{eq: ETH operator ansatz}, can be shown to be the spectral function associated with the operator $O$~\cite{DAlessio:2015qtq}.
It contains information on various scales associated with the operator thermalization.
Phenomenologically, it controls the transport properties when the operator $O$ corresponds to a conserved current density.
From Eq.~\eqref{eq: ETH operator ansatz}, it can be seen that $f^2_O(\bar{E}, \omega)$ is proportional to the off-diagonal MEs squared, i.e., $|\langle \bar{E}-\omega/2 | O |\bar{E}+\omega/2\rangle|^2$, averaged over a small energy window of width $\omega$, and the proportionality constant is $e^{-S(\bar{E})}$.
In this subsection, we first qualitatively analyze the dependence of $f_O(\bar{E}, \omega)$ on the local Hilbert space truncation, $\jmax$, for local operators in Fig.~\ref{fig: LSH_f_function jmax dependence} and extended operators in Fig.~\ref{fig: LSH_f_function_string jmax dependence}.
In the latter case, we also study their dependence on the operator length.
Finally, we study the behavior of $f_O(\bar{E}, \omega)$ at very large values of $\omega$ in Fig.~\ref{fig: f_function decay}.

In Fig.~\ref{fig: LSH_f_function jmax dependence} (a), we observe that $f_O$ for each local operator reaches a plateau at small frequencies while falling off fast at large frequencies.
The values of plateaus at small $\omega$ for $H_E/N$ and $H_I/N$ increase with $\jmax$, which can be understood from the fact that $H_E$ and $H_I$ depend on the gauge fields, while $H_M$ only contains fermion fields.
This suggests that one has to use even larger values of $\jmax$ to see a convergence in the $f_O$ function for $H_E/N$ and $H_I/N$.
Furthermore, since the sum of off-diagonal elements of $H_E$, $H_M$ and $H_I$ add up to zero in the energy eigenbasis, the increasing trends in the plateau values of $H_E/N$ and $H_I/N$ at $\omega\to0$ are not independent of each other.
We expect the plateau values to converge at sufficiently large values of $\jmax$ as the energy eigenstate within a fixed energy window receive increasingly smaller contributions from states with larger electric fields.
We confirm this convergence by using slightly larger parameter values, i.e., $g^2=0.64$ and $m=0.4$, as shown in Fig.~\ref{fig: LSH_f_function jmax dependence} (b). This choice of parameter values still falls within the quantum chaotic region in Fig.~\ref{fig: gap ratio param scan} (a).
At intermediate and large $\omega$ values, i.e., $|\omega|\gtrsim1$, we observe convergence with respect to $\jmax$ for all parameters and operators shown in Fig.~\ref{fig: LSH_f_function jmax dependence}. 

In Fig.~\ref{fig: LSH_f_function_string jmax dependence}, we study the $f_O$ function of extended operators.
ETH for non-local operators is not very well understood. Some analyses have already been done for spin systems~\cite{Bandyopadhyay:2022nrv}.
Two key observations for non-local string operators in~\cite{Bandyopadhyay:2022nrv} were the following:
\begin{enumerate}
    \item With increasing string size the zero frequency plateau height keeps decreasing. This is ascribed to formation of spectral gap characteristics of integrable systems \cite{Pandey:2020tru}.
    \item At low frequencies the $f_O$ function develops a {\em memory} peak indicative of a rapid late time oscillation in string auto-correlations computed in excited states. This contrasts with the late time behaviors of local operators, which decay rapidly, and hence is identified with memory.
\end{enumerate}
We find indications of both these effects in Fig.~\ref{fig: LSH_f_function_string jmax dependence}.
In Fig.~\ref{fig: LSH_f_function_string jmax dependence} (a), we plot the $f_O$ function of extended operator $S_5$ for increasing values of $\jmax$.
We note that for $j_{\mathrm{max}}=1$, there is no peak structure at intermediate $|\omega|$ between the exponential decrease and the plateau, but with increasing $j_{\mathrm{max}}$, we observe a (memory) peak emerging.
In Fig.~\ref{fig: LSH_f_function_string jmax dependence} (b), we compare the $f_O$ functions of different extended operators of varying lengths, $S_3$, $S_5$, and $S_7$, at a fixed value of $\jmax=3/2$. We observe that the $f_O$ function of $S_7$ has similar features as that of $S_5$, i.e., the lower plateau and the memory peak.
We note that, the behavior of $S_3$ as an extended operator is well observed in Figs.~\ref{fig: Lambda_n_measure} and~\ref{fig:semicircle_N12_jmax1.5_HM_S5}, however, with respect to the $f_O$ function, its features as an extended operator are less pronounced compared to $S_5$ and $S_7$ at truncation $j_{\rm max}=3/2$. However, it is expected that $S_3$ will show an increasingly pronounced peak at higher $j_{\rm max}$ values, similar to operator $S_5$ as depicted in Fig.~\ref{fig: LSH_f_function_string jmax dependence} (a).
The plateau value associated with $S_7$ is even smaller than that of $S_5$, indicating longer string operators are more integrable from the ETH perspective.
\begin{table}[h!]
    \renewcommand{\arraystretch}{1.5}
    \begin{center}
    \begin{tabular}{C{2cm}| C{2cm} C{2cm} C{2cm} C{2cm} C{2cm} }
     $O$ & $a$ & $b$ & $c$ & $\beta'$ & $\beta$ \\
    \hline
    \hline
    
    $H_E/N$ & 0.0004 & 0.0476 & 0.0808 & 4.9747 & 0.00016391 \\
    $S_3$ & 0.0029 & 0.0427 & 0.2323 & 3.7390 & 0.00016391 \\
    \hline
    \end{tabular}
    \end{center}\caption{Fitted parameters of the Lorentzian function in Eq.~\eqref{eq: Lorentian for spectral function at small omega} for the function $f_O(\bar{E},\omega)$ in the range $\omega\in[0,0.4]$ in Fig.~\ref{fig: f_function decay}, and the fitted value of $\beta'$ for $|\omega|\in[11,14]$ using the exponential function $e^{-\beta' |\omega|/4}$ are shown. For comparison, the calculated $\beta$ values are also listed. The errors in the fit values are small, and thus are not shown here.
    \label{tab: Lorentzian fit and beta}
    }
\end{table}

For local operators, after starting off with a plateau at $\omega =0$, and remaining featureless until some typical frequency, the $f_O$ function is expected to show a $1/\omega^2$ behavior for intermediate frequencies~\cite{Bertini_2021}. 
A phenomenological ansatz which captures both these regimes is a Drude-like (Lorentzian) function of the form~\cite{Schonle:2020grk}
\begin{align}
    f(|\omega|) = \frac{a}{\omega^2 + b} + c.
    \label{eq: Lorentian for spectral function at small omega}
\end{align}
On the other hand, at large frequencies the $f_O$ function is expected to decay exponentially and the decay rate is bounded from below by $e^{-\omega/4T}$~\cite{Murthy:2019fgs}.
Both of these limiting cases are shown in Fig.~\ref{fig: f_function decay} for $H_E/N$ and $S_3$ for $N=12$ and $\jmax=3/2$, the latter of which has been shown to behave like a local operator for $N=10$ with the same local Hilbert space truncation. 
We fitted the $f_O$ function in the range $|\omega|\in \left[0,0.4\right]$ by the Lorentzian function given in Eq.~\eqref{eq: Lorentian for spectral function at small omega} and the range
$|\omega|\in \left[11,14\right]$ by an exponential function $e^{-\beta'|\omega|/4}$, shown as solid green lines in the insets and dashed lines in Fig.~\ref{fig: f_function decay}, respectively.  Fitted parameter values are listed in Table~\ref{tab: Lorentzian fit and beta}.
The $\beta'$ values are $4.6086$ and $3.3299$ for $H_E/N$ and $S_3$, respectively.
These numbers are much bigger than the $\beta$ value given by the microcanonical or canonical temperature at which the $f_O$ function is calculated, which is very close to zero near the peak of the spectrum.
Thus we conclude that the decay rates at large frequency are consistent with the lower theoretical bound $\beta/4$. Furthermore, we find that the value of $\beta’$ decreases for increasing length of the string operator. This indicates a slower decay of the two-point correlation function at large frequency.

\section{Subsystem ETH: A preliminary study
\label{sec: Garrison-Grover}
}
The question for thermalization of all non-local operators that are supported on a given subsystem is identical to the question of thermalization of the subsystem~\cite{Garrison:2015lva}.
In this section we discuss some perspectives of subsystem ETH by studying how close a subsystem's reduced density matrix is to a thermal state, i.e.,
\begin{align}
    \rho_A(E) \equiv {\rm Tr}_{\bar{A}} (|E\rangle\langle E|) \stackrel{?}{=} \frac{1}{Z_A}e^{-\beta H_A},
    \label{eq: garrison-grover}
\end{align}

where the subscript $A$ denotes a subsystem with the complement $\bar{A}$, $|E\rangle$ is an eigenstate, $H_A$ stands for the subsystem Hamiltonian, $Z_A$ represents the subsystem thermal state normalization and the temperature $\beta$ is defined from the total system $H$ in the canonical sense
\begin{align}
    -\frac{\partial}{\partial \beta}\ln{\rm Tr}(e^{-\beta H}) =  E .
\end{align}
If Eq.~\eqref{eq: garrison-grover} is true, it follows that if we know the reduced density matrix $\rho_A$ of an eigenstate at a given energy $E_*$, which corresponds to the temperature $\beta_*$, we know the reduced density matrices for all eigenstates on the same subsystem, i.e.,
\begin{align}
    \rho_A(E) \propto [\rho_A(E_*)]^{\frac{\beta}{\beta_*}} .
    \label{eq: rho_gg}
\end{align}
One can then use Eq.~\eqref{eq: rho_gg} to compute observables that are defined on the subsystem $A$ and compare with the exact results obtained from the corresponding eigenstate.
This type of test was first done by Garrison and Grover~\cite{Garrison:2015lva} and it was shown to work well for the transverse field Ising model. Here we perform a similar analysis for the 1+1D SU(2) gauge theory.

A natural partition of the gauge theory lattice is to cut links. This has been used for $2+1$D SU(2) LGT without matter in the electric basis~\cite{Ebner:2024mee}, and it is particular useful for the LSH formulation since if the partition cuts through a vertex, it would destroy the gauge-invariant states constructed in the LSH basis.
A remaining Abelian constraint must be imposed for the cut links, which leads to decoupled Hilbert spaces on the subsystem that are specified by the cut boundary conditions. As a result, for gauge theory in general
\begin{align}
    \rho_A(E) = \bigoplus_{i} p_i(E) \, \rho_A(i,E),
\end{align}
where $i$ denotes different cut boundary conditions and $p_i$ is a classical distribution for them. 
We expect $p_i$ to depend on the eigenenergy $E$ and also to be correlated with the reduced density matrix $\rho_A(i,E)$, since both of them originate from the same eigenstate. This makes the Garrison-Grover test for LGTs dramatically different from spin models. With this reduced density matrix, one can calculate the entanglement entropy, which will yield a “classical” contribution to the entropy depending on the boundary links and a “quantum” contribution, also known as distillable entanglement entropy. The latter contribution can be extracted using local operations and classical communication (LOCC), but the former cannot~\cite{Soni:2015yga,VanAcoleyen:2015ccp}.

In the following, we assume that we know the eigenstate at energy $E_*$ completely and show observables ${\rm Tr}_A[O_A\rho_A(E)]$ at other energies calculated in three different ways:
\begin{subequations}
    \begin{align}
        \rho_A^{(1)}(E) &\propto [\rho_A(E_*)]^{\frac{\beta}{\beta_*}} , \label{eq: Garrison Grover original}\\
        \rho_A^{(2)}(E) &= \bigoplus_{i} p_i(E_*) \frac{1}{Z_A(i,\beta)}[\rho_A(i,E_*)]^{\frac{\beta}{\beta_*}} ,\label{eq: Garrison Grover unadjusted block}\\
        \rho_A^{(3)}(E) &= \bigoplus_{i} p_i(E) \frac{1}{Z_A(i,\beta)}[\rho_A(i,E_*)]^{\frac{\beta}{\beta_*}}\label{eq: Garrison Grover adjusted block} .
    \end{align}
    \label{eq: Garrison Grover all three}%
\end{subequations}
\begin{figure*}[t!]
    \centering
    \includegraphics[scale=1.0]{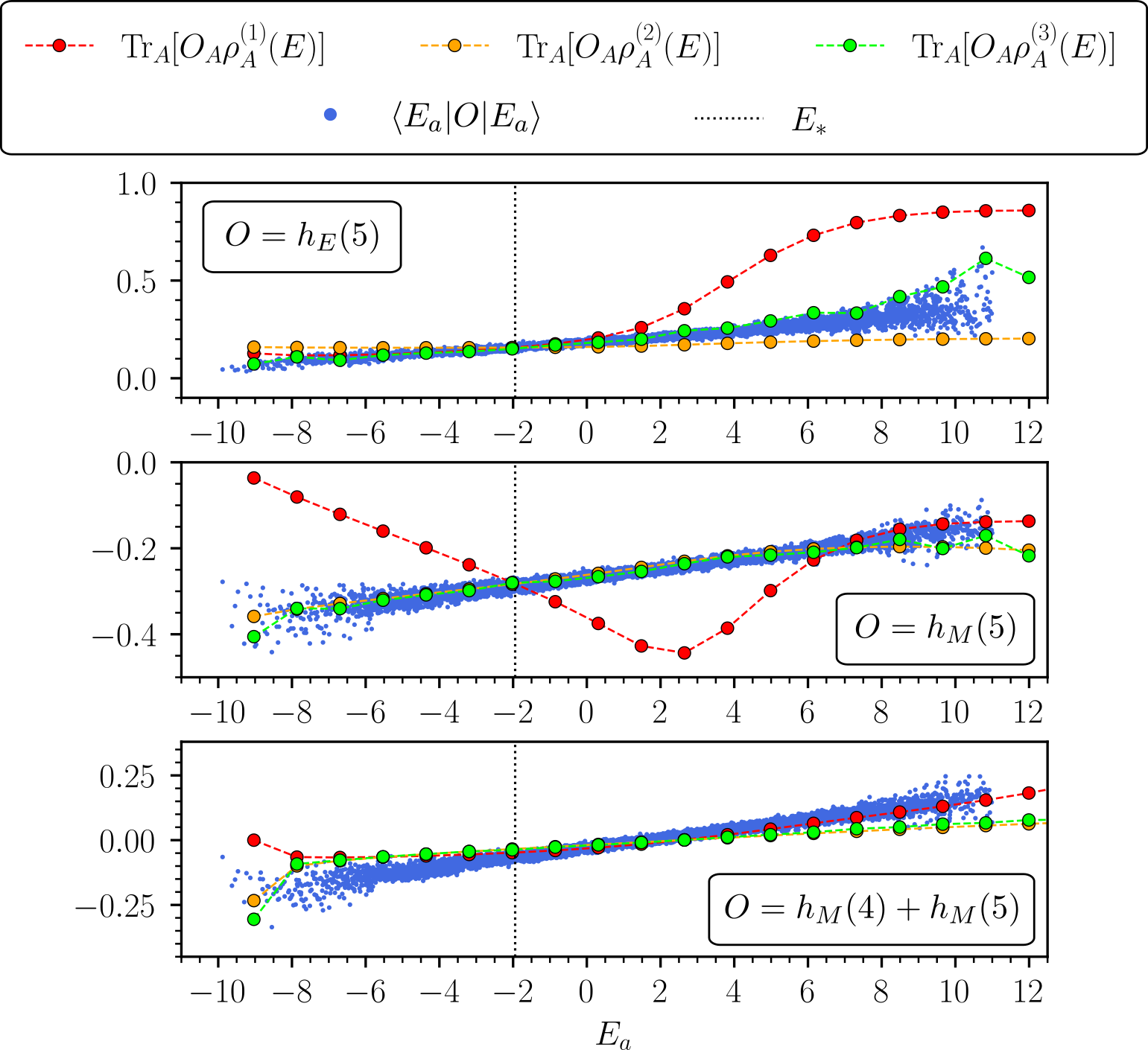}
    \caption{We compare the predictions from different $\rho^{(i)}_A(E)$ used in the Garrison-Grover procedure in Eq.~\eqref{eq: Garrison Grover all three} for the microcanonical values of an operator that dominate its diagonal MEs according to ETH.
    The predictions from $\rho^{(i)}_A(E)$ for $i=1$, 2, and 3 are shown with red, yellow, and green dots, respectively. The diagonal MEs are shown with blue dots. The dotted black line indicates the energy $E_*$ of the state 
    chosen in Eq.~\eqref{eq: Garrison Grover all three}. We considered three operators $h_E(5)$, $h_M(5)$, and $h_M(4)+h_M(5)$ in the top, middle and bottom panel, respectively. In all three cases,  the results from $\rho^{(1)}_A$ show significant disagreement with the diagonal MEs. However, $\rho^{(2)}_A$ and $\rho^{(3)}_A$ predictions agree well with the diagonal MEs and with each other for $h_M$ but show dependencies for $h_E$. In the latter case $\rho^{(3)}_A$ aligns closely with the diagonal MEs. 
    \label{fig: gg_test}
    }
\end{figure*}

The first method in Eq.~\eqref{eq: Garrison Grover original} is the original Garrison-Grover test.
The second method in Eq.~\eqref{eq: Garrison Grover unadjusted block} takes into account the direct sum structure in partition of a gauge theory system but only applies the $\beta$ scaling as in the original test to $\rho_A(i,E)$.
However, the classical distribution $p_i(E_*)$ is taken to be that at the given energy $E_*$ and does not change with $E$ or equivalent $\beta$. The boundary-dependent normalization $Z_A(i,\beta)$ was introduced such that
\begin{align}
    \frac{1}{Z_A(i,\beta)} {\rm Tr}\Big\{[\rho_A(i,E_*)]^{\frac{\beta}{\beta_*}}\Big\} = 1.
\end{align}
In the third method in Eq.~\eqref{eq: Garrison Grover adjusted block}, the energy dependence in $p_i$ is accounted for by calculating them for each eigenstate at the energy $|E\rangle$. In other words, in Eqs.~\eqref{eq: Garrison Grover original} and~\eqref{eq: Garrison Grover unadjusted block}, only the state at the energy $E_*$ is used to obtain states at other energies while in Eq.~\eqref{eq: Garrison Grover adjusted block}, eigenstates at other energies are also used. 

Our results are shown in Fig.~\ref{fig: gg_test} for a subsystem of length 3 on a $N=12$ lattice with $g^2=0.5$ and $m=0.5$. In this calculation open boundary conditions are used with $\mathcal{B}=-6$ and $\Delta \mathcal{Q}=0$, defined in Eqs.~\eqref{eq: baryon number B definition} and~\eqref{eq: Delta Q sq definition}, respectively, and we use the full Hilbert space on this lattice.
The subsystem is from site $4$ to $6$. Our choice of the subsystem size is based on the fact that we want a bath that is larger than half of the system while reducing boundary effects. The energy $E_*$ of the given state is chosen near the middle of the spectrum, shown by the dotted black line in Fig.~\ref{fig: gg_test}, which corresponds to $\beta_*=0.3$. 
The local electric energy operator is calculated using Eq.~\eqref{eq: hE in LSH} for $x=5$, while the fermion mass energy is computed using Eq.~\eqref{eq: hM in LSH} for $x=4$, $5$.
As can be seen, the observables calculated using $\rho^{(1)}_A$ and $\rho^{(2)}_A$ deviate from the exact results for $h_E(5)$ as the energy moves away from $E_*$.
However, for operators independent of gauge link values, i.e., $h_M(x)$, $\rho^{(2)}_A$ gives good agreement with the numerical values of the diagonal MEs.
On the other hand, the results from $\rho^{(3)}_A$ follow closely the diagonal MEs values for a wider range of energies and both $h_E$ and $h_M$ operators.
This strongly indicates that the energy dependence of the classical distribution $p_i$ must be taken into account, and it is not given by the simple $\beta$ scaling as in the Garrison-Grover procedure originally proposed in~\cite{Garrison:2015lva}.
This highlights a unique feature of gauge theory that makes it distinct from simple spin models.
Finally, the analytic dependence of $p_i$ on $E$ might provide interesting insights into the Hilbert space structure of LGTs.
We leave further exploration to gain a better understanding of $\rho_A(E)$ for LGT subsystems to future study.

\section{Conclusions
\label{sec: conclusions}}

We investigated the (1+1)-dimensional SU(2) lattice gauge theory (LGT) coupled to dynamical fermions in the loop-string-hadron (LSH) framework with respect to the validity of the eigenstate thermalization hypothesis (ETH).
In the ergodic regime of the parameter space, we found that all considered operators, local and non-local, approach the ETH behavior for sufficiently small energy windows. In detail, we first identified the region of $g$ and $m$ parameter space for which the energy spectrum of the momentum-zero, charge- and parity-even sector obeys the Gaussian orthogonal ensemble (GOE) predictions.
We achieved this via calculating the mean restricted gap ratio, $\langle r \rangle$, and the spectral form factor (SFF).
The latter also contained the information of the time scale at which universal random matrix theory (RMT) behavior occurs at the spectral level. 
For a specific set of parameters in the ergodic region, we calculated matrix elements (MEs) of local and non-local operators in the eigenbasis and showed that a smooth microcanonical value emerges in the thermodynamic limit.
We confirmed the exponential suppression in the variances of both diagonal and off-diagonal MEs with the system size, namely $e^{-S(E)}$, and found that the averaged entropy density obtained from the exponential decrease rate is independent of the operators.
Next, we used different sensitive GOE measures to test the randomness of the correction in the MEs to the microcanonical values. 
We found that $\langle r \rangle$ calculated from the eigenvalues of a band matrix was the least sensitive measure, as its value agreed with the GOE value in the whole energy window considered ($\Delta E=0.75$). 
The $\Sigma$ and $\Gamma$ measures defined in Eqs.~\eqref{eq: Sigma definition} and~\eqref{eq: Gamma measure definition}, respectively, are more informative but they are only sensitive to absolute values of the MEs and miss important sign correlations. 
We found these two measures for all operators approach the GOE predictions at late time, i.e., in smaller energy windows. 
On the other hand, the $\Lambda_n(E,t)$ measure in Eq.~\eqref{eq: Lambda n definition} and the semicircle distribution of operator band matrix eigenvalues are sensitive to the sign correlations and need much longer time, i.e. even smaller energy windows, to show the GOE behavior for all the operators.
It is worth highlighting that not only the local but also the non-local operators show ETH behavior at late time. 
Moreover, not all operators approach the GOE prediction for the same measure at the same time. 
Furthermore, we extracted the function $f_O$ from the off-diagonal MEs. 
We found that its small-frequency part can be well described by a Lorentzian shape and its large-frequency behavior is consistent with the theoretical upper bound.
Finally, we studied some aspects of subsystem ETH by performing the Garrison-Grover test and showed that the reduced density matrix of a gauge theory system is more complicated than that of chaotic Ising-type models.

The theory of subsystem thermalization plays an important role in understanding the behavior of gauge theories in non-equilibrium dynamics such as the expansions in the early Universe and heavy ion collisions.
Thus, further studying subsystem thermalization in the LGT considered here by investigating the entanglement entropy and analyzing the structure of the subsystem density matrix that is unique to gauge theories could provide novel information that allows to better understand non-equilibrium dynamics in gauge theories.
This would also improve simulation of dynamical phenomena in gauge theories, that are presently still limited to theories that are much simpler than QCD. 
Combining this better understanding with emerging new tools, like quantum computing and tensor network methods can, furthermore, be expected to greatly speed up all such developments.
In the 1+1D non-Abelian LGT studied here, the gauge fields participate only through the electric energy and their interactions with the dynamical fermions.
However, in higher spatial dimensions, the gauge fields also contribute to the total energy via the magnetic term.
This allows for non-trivial dynamics involving only the self-interacting non-Abelian gauge fields in absence of matter degrees of freedom. 
The ETH ansatz for some operators in such theories has been verified recently~\cite{Yao:2023pht,Ebner:2023ixq}, however, inclusion of dynamical matter in a theory with dynamical gauge fields and its effects on quantum chaos and ETH operator structure remain to be explored. 
In this respect it is also interesting to note that Ref.~\cite{Murthy:2022dao} recently formulated a version of the ETH in the presence of a {\em global} non-Abelian symmetry wherein one factors out the Clebsch-Gordon coefficients form the operator MEs and analyses the remaining factors for ETH. 
Another kind of generalization of ETH in the form of allowing for non-Gaussianly correlated off-diagonal MEs has also been explored~\cite{Foini:2018sdb, Murthy:2019fgs}. 
This latter generalization is a requirement for chaotic systems to reproduce Lyapunov behavior in out-of-time-ordered correlation functions. 
Furthermore, some recent progress has also been made in understanding thermalization of non-Abelian LGTs from other perspectives~\cite{Guin:2025lpy,Pandey:2024goi,pandey2025thermalsu2latticegauge}.
It would be interesting to explore all these developments in the context of LGTs.
Finally, getting closer to understanding the non-equilibrium dynamics of the Standard Model of particle physics requires similar analyses to be performed with the SU(3) group, which we plan to do in the future.

\acknowledgements{}
This work originated from the discussions at ``Thermalization, from Cold Atoms to Hot Quantum Chromodynamics" workshop (September 2023) held at the InQubator for Quantum Simulation (IQuS)\footnote{\url{https://iqus.uw.edu/}} hosted by the Institute for Nuclear Theory and also at the ``Susegad Symposium on Physics with Quantum Technology (SSPQT): NISQ-era and Beyond" (January 2024) at BITS Pilani, Goa campus, sponsored by the cross-discipline research fund (C1/23/185) from BITS-Pilani.
The authors gratefully acknowledge the scientific support and HPC resources provided by the Erlangen National High Performance Computing Center (NHR@FAU) of the Friedrich-Alexander-Universität Erlangen-Nürnberg (FAU) under the NHR project ID b172da. 
NHR funding is provided by federal and Bavarian state authorities. NHR@FAU hardware is partially funded by the German Research Foundation (DFG) – 440719683.
This work was enabled, in part, by the use of advanced computational, storage and networking infrastructure provided by the Hyak supercomputer system at the University of Washington.
The authors would like to thank Jesse Stryker for helpful discussions in the initial stage of this work. L.E. thanks Leonhard Schmotzer for informative discussions about the spectral form factor.
S.K. acknowledges valuable discussions with Michael Winer.

L.E.~acknowledges funding by the Max Planck Society, the Deutsche Forschungsgemeinschaft (DFG, German Research Foundation) under Germany’s Excellence Strategy – EXC-2111 – 390814868, and the European Research Council (ERC) under the European Union’s Horizon Europe research and innovation program (Grant Agreement No.~101165667)—ERC Starting Grant QuSiGauge.   
Views and opinions expressed are however those of the author(s) only and do not necessarily reflect those of the European Union or the European Research Council Executive Agency. Neither the European Union nor the granting authority can be held responsible for them. 
This work is part of the Quantum Computing for High-Energy Physics (QC4HEP) working group.
I.R. is supported by the  OPERA award (FR/SCM/11-Dec-2020/PHY) from BITS-Pilani, the Start-up Research Grant (SRG/2022/000972) and Core-Research Grant (CRG/2022/007312) from ANRF(SERB), India and the cross-discipline research fund (C1/23/185) from BITS-Pilani.
S.K. and X.Y. are supported by the U.S. Department of Energy, Office of Science, Office of Nuclear Physics, IQuS under Award Number DOE (NP) Award DE-SC0020970 via the program on Quantum Horizons: QIS Research and Innovation for Nuclear Science\footnote{\url{https://science.osti.gov/np/Research/Quantum-Information-Science}}.
S.K. also acknowledges the U.S. Department of Energy QuantISED program through the theory consortium ``Intersections of QIS and Theoretical Particle Physics'' at Fermilab (Fermilab subcontract no. 666484).
This work was also supported, in part, through the Department of Physics and the College of Arts and Sciences at the University of Washington. 

\appendix
\section{Symmetry transformations of field operators
\label{app: symmetry on operators}}
In this appendix, we list the symmetry transformation properties of various field operations as used in Sec.~\ref{subsubsec: symmetries}. 
We provide here the definitions of the symmetry transformations for the field operators that have been depicted in Fig.~\ref{fig: KS and Schwinger dofs}. 
These relations can be used along with the definitions of the LSH operators in Table~\ref{tab: LSH operators} to verify that the Hamiltonian in Eq.~\eqref{eq: complete Hamiltonian in KS} is invariant under these transformations in both, the KS and the LSH formulations.

The spatial translation operator ${\mathcal{T}}_2$ that shifts all the degrees of freedom by two staggered sites has an obvious realization for the field operators:
\begin{subequations}
    \begin{align}
       {\mathcal{T}}^{-1}_2 \, \psi_\alpha(x) \, {\mathcal{T}}_2 &= \psi_\alpha(x+2), \label{subeq: translation on psi}\\
        {\mathcal{T}}^{-1}_2  \, E^{\rm a}_s(x) \, {\mathcal{T}}_2&= E^{\rm a}_s(x+2), \label{subeq: translation on Es}\\ 
        {\mathcal{T}}^{-1}_2 \, U(x,x+1) \, {\mathcal{T}}_2 &= U(x+2,x+3), \label{subeq: translation on U}\\
        {\mathcal{T}}^{-1}_2 a_\alpha(s,x){\mathcal{T}}_2 &= a_\alpha(s,x+2), \label{subeq: translation on as} %
    \end{align}
\end{subequations}
where the last equation is the transformation rule of the Schwinger boson that equivalently achieves the transformations shown in the second and third lines.
For lattice with PBCs, it is implicit that $x+2$ is taken modulo $N$, so the Hamiltonian is invariant under this transformation.

The parity operator ${\mathcal{P}}$ reflects the lattice about the axis chosen to pass through the site at $N/2-1$.
Under the parity operation, the left and right electric field operators transform into each other and the link operator to its Hermitian conjugate.
Mathematically, the operator ${\mathcal{P}}$ gives
\begin{subequations}
    \begin{align}
       {\mathcal{P}}^{-1} \, \psi_\alpha(x) \, {\mathcal{P}} &= \psi_\alpha(x-2i), \label{subeq: parity on psi}\\
        {\mathcal{P}}^{-1} \, E^{\rm a}_s(x) \, {\mathcal{P}} &= E^{\rm a}_{s'}(x-2i), \label{subeq: parity on Es}\\ 
        {\mathcal{P}}^{-1} \, U(x,x+1) \, {\mathcal{P}} &= U^\dagger(x-2i-1,x-2i), \label{subeq: parity on U}
    \end{align}
    \label{eq: parity on psi Es U}%
\end{subequations}
where $s' = R$ $(L)$ if $s=L$ $(R)$, and $x = N/2 - 1 + i$. 
The Hamiltonian is invariant under ${\mathcal{P}}$ for a lattice with PBCs.
Thus, it is understood that all position numbers are taken modulo $N$.
It can be seen from Eqs.~\eqref{eq: U in UL and UR} and~\eqref{eq: UL and UR in schwinger bosons} that an equivalent transformation of the Schwinger boson operators is given by
\begin{subequations}
    \begin{align}
        {\mathcal{P}}^{-1}\, a_\alpha(L,x) \,{\mathcal{P}} &= (-1)^x\, a_\alpha(R,x-2i),\label{subeq: parity on aL}\\
        {\mathcal{P}}^{-1} \, a_\alpha(R,x) \, {\mathcal{P}} &= (-1)^x\, a_\alpha(L,x-2i).\label{subeq: parity on aR}
    \end{align}
    \label{eq: parity on aL and aR}%
\end{subequations}
\begin{figure}[t]
    \includegraphics[scale=1.0]{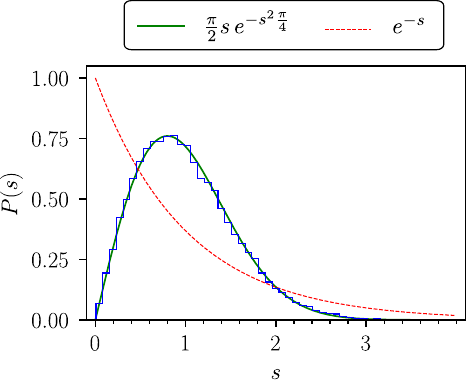}
    \caption{Histogram of probability density (blue) of the unfolded level spacings $s$ compared to the GOE (green solid) and the Poisson (red dashed) predictions. 
    The system is $N=14$ with the parameters $j_{\mathrm{max}}=1/2$, $m=0.25$ and $g^2=0.25$.
    The agreement with the GOE prediction indicates that the LGT at these Hamiltonian parameters is quantum chaotic.
    \label{fig: level spacing}}
\end{figure}

Finally, the charge conjugation transformation changes particles into anti-particles.
Thus, its realization ${\mathcal{C}}$ on the staggered lattice is given by
\begin{subequations}
    \begin{align}
       {\mathcal{C}}^{-1} \, \psi_\alpha(x) \, {\mathcal{C}} &= (-1)^x (-i) \psi^\dagger_\alpha(x+1), \label{subeq: CC on psi}\\
        {\mathcal{C}}^{-1}  \, E^{\rm a}_s(x) \, {\mathcal{C}}&= \eta_{\rm a2} \, E^{\rm a}_s(x+1), \label{subeq: CC on Es}\\ 
        {\mathcal{C}}^{-1} \, U(x,x+1) \, {\mathcal{C}} &= U^{*}(x+1,x+2), \label{subeq: CC on U}
    \end{align}
\end{subequations}
where $\eta_{\rm a2} = -1$ (+1) if ${\rm a}\neq 2$ ($\rm a=2$), and $*$ denotes the complex conjugate. In terms of the Schwinger bosons, the charge conjugation transformation is given by
\begin{subequations}
    \begin{align}
        {\mathcal{C}}^{-1}\, a_\alpha(L,x) \,{\mathcal{C}} &= i\, \epsilon_{\alpha\beta} \, a_\beta(L,x+1),\label{subeq: CC on aL}\\
        {\mathcal{C}}^{-1}\, a_\alpha(R,x) \,{\mathcal{C}} &=-i\, \epsilon_{\alpha\beta} \, a_\beta(R,x+1). \label{subeq: CC on aR}
    \end{align}
    \label{eq: CC on aL and aR}%
\end{subequations}
Equations~\eqref{eq: parity on psi Es U}-\eqref{eq: CC on aL and aR} define the equivalent transformations on the LSH operators that leave the Hamiltonian invariant.
Note that, the Hamiltonian is invariant only when it is considered in the physical Hilbert space, that is, the LSH states satisfy the Abelian Gauss's law condition in Eq.~\eqref{eq: Abelian gauss law constraint on LSH qunatum numbers}, which assures that the square root factors on each end in Eq.~\eqref{eq: hI in LSH} are equal.

\section{Level spacing statistics and unfolding
\label{app: level statistics}
}
Here we study the distribution of level spacings in the LSH Hamiltonian. 
For Hamiltonians with different parameters, the range of the energy spectrum can be vastly different. 
Furthermore, the mean of the level spacings can be energy dependent.
Therefore, in order to make a well-defined comparison with the GOE whose mean level spacing is one,
we unfold the spectrum i.e., normalize the level spacing $\delta_\alpha$ by taking its ratio with the local average level spacings $\Delta(E)=\frac{1}{\rho(E)}$ such that the mean level spacing is one at all energies~\cite{Brody:1981cx,Jackson:2000zr}.
\begin{table}[b!]
    \renewcommand{\arraystretch}{1.3}
    \begin{center}
    \begin{tabular}{C{1.8cm} C{2.5cm}|C{1.8cm} C{2.5cm}}
     Coefficient & Fit Value & Coefficient & Fit Value\\
    \hline
    \hline
        $a_0$ & $2.22\times10^{-4}$ & $a_1$ & $- 1.09\times10^{-5}$\\
        $a_2$ & $7.33\times10^{-6}$ & $a_3$ & $- 2.47\times10^{-7}$\\
        $a_4$ & $1.21\times10^{-7}$ & $a_5$  & $- 7.68\times10^{-9}$\\  
        $a_6$ & $1.36\times10^{-9}$ & $a_7$ & $3.78\times10^{-11}$\\
        $a_8$ & $2.53\times10^{-12}$  &  $a_9$ & $- 1.11\times10^{-12}$\\
        $a_{10}$  & $1.66\times10^{-13}$ & &\\
    \hline
    \end{tabular}
    \end{center}\caption{Fitted coefficients of the tenth order polynomial $p_{10}(E)= \sum_{i=0}^{10} a_i E^i$ used for obtaining the unfolding in Fig.~\ref{fig: level spacing}. Here, $a_i$ denotes the coefficient of the monomial of order $i$, and the fitting is performed for the energy spectrum of the Hamiltonian with $N=14$, $\jmax=1/2$, $m=0.25$ and $g^2=0.25$.
    \label{tab:p10}}
\end{table}

We do this by fitting a polynomial of order 10, i.e., $p_{10}(E)$ to the raw level spacing at the middle point $\delta_a(\frac{E_a+E_{a+1}}{2})$. The fitted coefficients of the 10th-order polynomial are listed in Table~\ref{tab:p10}. 
The unfolded energy gaps are then $s_{a}\equiv\delta_{a}/ p_{10}(E_{a})$.
The distribution of the unfolded level spacings for a $N=14$ system with $j_{\mathrm{max}}=1/2$, $m=0.25$ and $g^2=0.25$ is shown in Fig.~\ref{fig: level spacing} and compared to the GOE prediction for a $2 \times 2$ matrix given by
\begin{equation}
     P(s)=\frac{\pi}{2}s\,e^{-s^2\frac{\pi}{4}}.
    \label{eq: GOE distribution}
\end{equation}
We find that they agree very well, indicating that the theory at these parameters is non-integrable and quantum ergodic.

\section{Spectrum dependence on  the local bosonic Hilbert space truncation
\label{app: spectrum jmax}
}
\begin{figure*}[t!]
    \centering
  \includegraphics[scale=1.0]{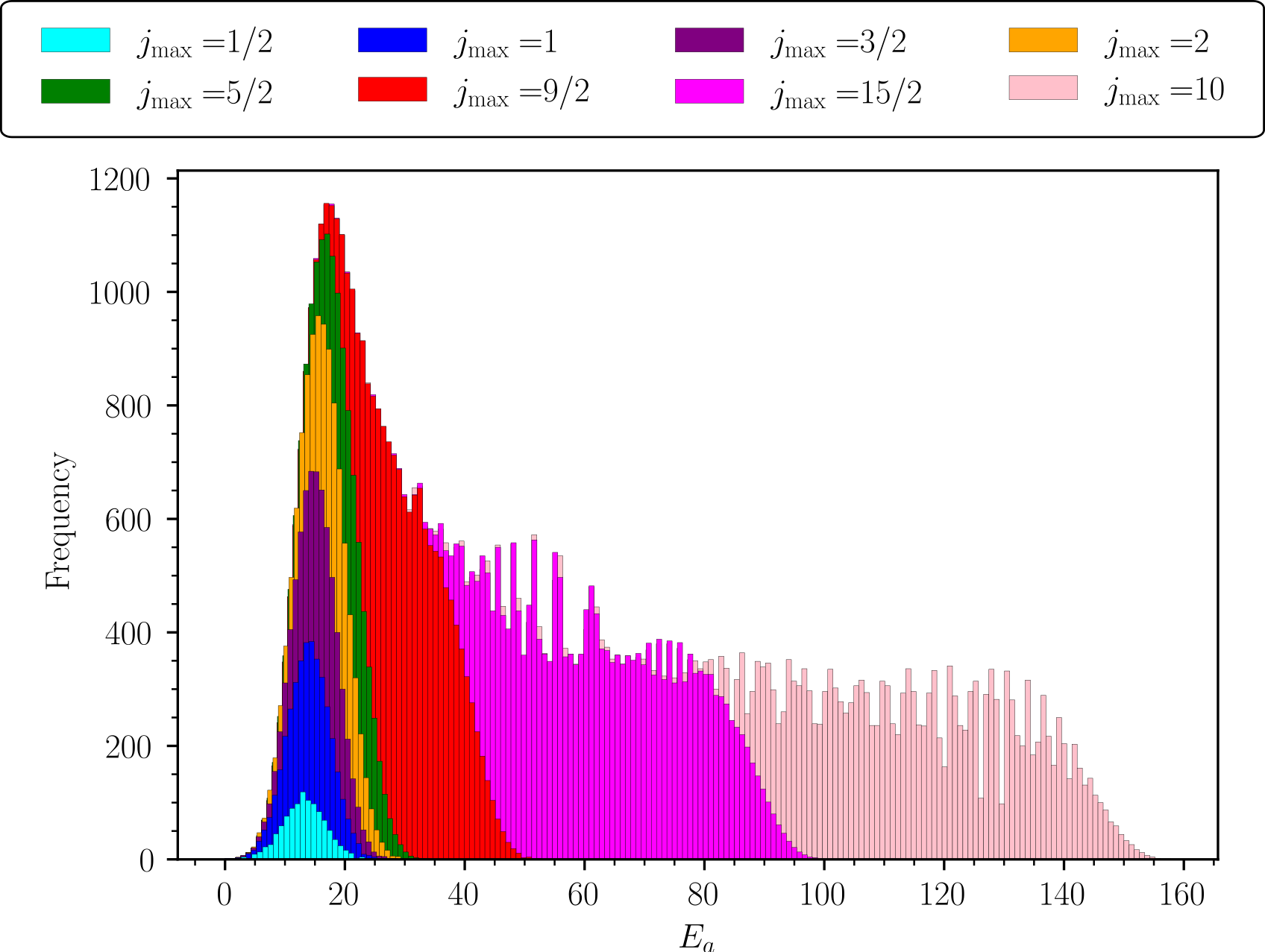}
    \caption{Frequency histograms of the energy spectra for different $j_{\mathrm{max}}$ for a $N=10$ periodic Hamiltonian lattice system with parameters $g^2=0.25$ and $m=0.25$. 
    We consider the same symmetry sector as in Sec.~\ref{sec: Results}, i.e. translational invariant states with $\mathcal{B}=0$ that are even under parity and charge conjugation.
    The size of each bin is fixed to 0.87 for all cases. The convergence is observed for energies $E \leq 15$ at $\jmax= 5/2$. This is seen from the fact that newly added states for higher $\jmax$ have larger energies.
    \label{fig: jmax depedence of density of states}
    }
\end{figure*}

In this appendix, we study the effects of $j_{\mathrm{max}}$ (defined in Eq.~\eqref{eq: jmax truncation on Ns}) on the density of states, $\rho(E)$, and the quantum ergodicity indicator $\langle r\rangle$.
In Fig.~\ref{fig: jmax depedence of density of states}, we show $\rho(E)$ for various values of $\jmax\leq 10$ for a Hamiltonian system with PBCs, the lattice size $N=10$, and parameters $g^2=0.25$ and $m=0.25$. The dimensions of the respective Hilbert spaces are listed in Table~\ref{tab: Hilbert space dimension for various jmax}. 
First, we note that the low energy spectrum for $E<15$ converges at $\jmax=5/2$, indicating that any further increase in $\jmax$ does not affect the physics in this region anymore.
\begin{table}[t!]
    \renewcommand{\arraystretch}{1.5}
    \begin{center}
    \begin{tabular}{C{1.5cm}| C{1.3cm} C{1.3cm} C{1.3cm} C{1.3cm} C{1.3cm} C{1.3cm} C{1.3cm} C{1.3cm} }
     $\jmax$ & $1/2$ & $1$ & $3/2$ & $2$ &  $5/2$ & $9/2$ & $15/2$ & $10$ \\
     \hline
     \hline
    $\mathcal{D}$ & 1066 & 3960 & 7421 & 10928 & 14436 & 28468 & 49516 & 67056\\
    \hline
    \end{tabular}
    \end{center}\caption{ The dimensions of the Hilbert spaces $(\mathcal{D})$ for each truncation $\jmax$ in Fig.~\ref{fig: jmax depedence of density of states} are provided in this table.
    \label{tab: Hilbert space dimension for various jmax}
    }
\end{table}
On the other hand, we observe the lengthening of the tail in the high energy region of the spectrum with increasing $\jmax$ since the newly added states are more energetic.
This is due to the fact that these states correspond to increasing values of the background electric flux around the lattice.
As a result, the quantum ergodicity may be affected in that region.
We observe this in the indicator $\langle r \rangle$: For $j_{\mathrm{max}}=10$, we obtain $\langle r \rangle=0.5316$ for states with energies in the range $0\leq E\leq 50$, while $\langle r \rangle=0.5169$ for $50\leq E\leq 158$.
The former is very close to the GOE value $\langle r \rangle=0.5307$, while the latter shows some deviation away from it, but still far from the Poisson value $\langle r \rangle=0.3863$.
We emphasize that all ETH indicators in this work were analyzed in the energy region where $\langle r \rangle$ is within roughly 5\% of the GOE prediction.

%
\end{document}